\documentclass[12pt,preprint]{aastex}
\usepackage{color}

\shorttitle{Aperture Infrared Polarimetry of M42}
\shortauthors{Kusakabe et al.}

\begin{document}

\title{Near-Infrared Imaging Polarimetry of M42: Aperture Polarimetry of Point-like Sources}

\author{Nobuhiko \textsc{KUSAKABE},\altaffilmark{1}  
        Motohide \textsc{TAMURA},\altaffilmark{1,2}
        Ryo \textsc{KANDORI},\altaffilmark{2}        
        }
\author{Jun \textsc{HASHIMOTO},\altaffilmark{1,3}
Yasushi \textsc{NAKAJIMA},\altaffilmark{1} }
\author{Tetsuya \textsc{NAGATA},\altaffilmark{4} 
        Takahiro \textsc{NAGAYAMA},\altaffilmark{4} 
        Jim \textsc{HOUGH},\altaffilmark{5}
		Phil \textsc{LUCAS}\altaffilmark{5}}

\altaffiltext{1}{Graduate University of Advanced Science, 2-21-1 Osawa, Mitaka, Tokyo 181-8588}
\altaffiltext{2}{National Astronomical Observatory, 2-21-1 Osawa, Mitaka, Tokyo 181-8588}
\altaffiltext{3}{Department of Physics, Tokyo University of Science, 1-3, Kagurazaka, Shinjuku-ku, Tokyo 162-8601}
\altaffiltext{4}{Department of Astronomy, Kyoto University, Sakyo-ku, Kyoto 606-8502}
\altaffiltext{5}{Centre for Astrophysics Research, University of Hertfordshire, Hatfield, Herts AL10 9AB, UK}
\email{kusakabe@optik.mtk.nao.ac.jp}



\begin{abstract}
We have conducted aperture polarimetry of $\sim$500 stars of the Orion Nebula Cluster (ONC)
in M42 based on our wide-field ({$\sim$}8$'${$\times$}8$'$) $JHKs$ band polarimetry.
 Most of the near-infrared (NIR) polarizations are dichroic, 
with position angles of polarization agreeing, both globally and locally, 
with previous far-infrared (FIR) and submillimeter observations, having 
taken into account the 90$^\circ $ difference in angles between dichroic absorption and emission.
This is consistent with the idea that both NIR dichroic polarizations
and FIR/submillimeter thermal polarizations trace the magnetic fields
in the OMC-1 region.
The magnetic fields inferred from these observations
show a pinch at scales less than 0.5 pc with a centroid
near IRc2.
The hourglass-shaped magnetic field pattern
is explained by the models in which the magnetic field
lines are dragged along with the contracting gas and then
wound up by rotation in a disk.
The highly polarized region to the northwest of IRc2 and
the low-polarized region near the bright bar
are also common among NIR and FIR/submillimeter data,
although a few regions of discrepancy exist.

We have also discerned {$\sim$}50 possible highly polarized sources
whose polarizations are more likely to be intrinsic
rather than dichroic.
Their polarization efficiencies ($P(H)/A(H)$) are too large
to be explained by the interstellar polarization.
These include 10 young brown dwarfs that
suggest a higher polarization efficiency,
which may present geometrical evidence for
(unresolved) circumstellar structures
around young brown dwarfs.

\end{abstract}

\section{Introduction}
Magnetic fields are believed to play an important role in the evolution of molecular clouds, 
from their large-scale structures to dense cores, protostar envelopes, and protoplanetary disks.
Magnetic fields can be measured by observing polarizations of electromagnetic waves.
Dichroic polarizations are believed to be caused by spinning non-spherical dust grains becoming aligned with their short axis precessing around the direction of the local magnetic field.  In absorption this produces polarization parallel to the magnetic field and perpendicular to the magnetic field in emission 
(e.g., Weintraub et al. 2000).
Optical stellar polarimetry is useful for revealing magnetic field 
structures in the periphery of molecular clouds 
(e.g., Appenzeller 1974; Vrba et al. 1976; Moneti et al. 1984; Goodman et al. 1990).

However, because of heavy dust extinction, optical polarimetry is not always 
a good tracer of the magnetic fields inside molecular clouds.
Since the pioneering work by Vrba et al. (1976) and Wilking et al. (1979), 
near-infrared (NIR) stellar polarimetry in star-forming regions has been a powerful tool 
for tracing magnetic field structures in molecular clouds 
(Tamura et al. 1987; Sato et al. 1988; Tamura et al. 1988; Klebe \& Jones 1990).
The stars could be either background stars or even embedded sources if they are not 
associated with circumstellar structures (i.e., have no intrinsic polarization).
Although arguments have been made against the usefulness of NIR polarimetry 
in tracing magnetic fields inside molecular clouds and despite a suggestion that 
the dust grains are not well aligned in the dense regions (Goodman et al. 1995), 
recent submillimeter polarimetry has clearly shown aligned dust grains, 
even in cold regions such as starless globules and cores (Kirk, Ward-Thompson, \& Crutcher 2006).
With the advent of wide-field NIR polarimetric capability (Tamura et al. 1996), 
reevaluating the utility of NIR stellar polarimetry is warranted because 
the wide field allows simultaneous polarimetry of many stars in the field of view.
The Orion Nebula Cluster (ONC), or Trapezium cluster, is an ideal site for such a study.
The ONC is one of the nearest (450 pc), massive, star-forming regions to the Sun and the most populous 
young cluster within 2 kpc, composed of some 3500 young, low-mass stars (O'Dell 2001).\\

The magnetic field structure of the Orion region was most extensively
studied by observing linearly polarized thermal emission from aligned dust grains.
Houde et al. (2004) and Schleuning (1998) showed at 350 $\mu$m and 100$\mu$m, respectively, 
that the magnetic field in OMC-1 is generally oriented northwest--southeast, 
with the field pinched along the northeast--southwest direction on a scale of several arcmin. 
The polarization percentage is low at the location of the Becklin--Neugebauer object and Kleinmann--Low nebula (BN/KL) compared to elsewhere, 
but that may be due to a small-scale variation that is undetectable in their 12$''$ - 35$''$ resolution maps.
See Cudlip et al. 1982, Hildebrand et al. 1984,
Dragovan 1986, Barvainis et al. 1988, Gonatas et al. 1990,
Leach et al. 1991, and
Rao et al. 1998 for earlier or other millimeter and submillimeter studies in the Orion region.

Imaging polarimetry of M42 has been conducted either only at optical wavelengths (Pallister et al. 1977) 
or toward a small region near IRc2 or BN at near-infrared wavelengths (Minchin et al. 1991; 
Jiang et al. 2005; Simpson et al. 2006).
Hough et al. (1986), Burton et al. (1991), and Chrysostomou et al. (1994) measured 
the polarization of the H$_2$~v~=~1-0~S(1) line at $2.12\mu$m around IRc2. 
They attributed the polarization in the vicinity of IRc2 to dichroic absorption, and in particular, 
the twist in the polarization vectors 5$''$ east of IRc2 to a twist in the magnetic field direction.
At NIR wavelengths, the other IRc sources (IRc2 is not readily visible at 2~$\mu$m, 
probably because it is behind the edge of the hot core) are all strongly polarized, 
with centrosymmetric polarization vectors indicating scattered light; 
that is, the IRc sources are illuminated by either BN (IRc1) or some star 
in the vicinity of IRc2 (Werner et al. 1983; Minchin et al. 1991; Chrysostomou et al. 1994).

The high degree of polarization of BN, which is proportional to the extinction 
with the FIR-inferred magnetic field directions, has been used to deduce 
that BN is polarized by dichroic absorption at all NIR and MIR wavelengths 
where it has been measured 
(e.g., Lee \& Draine 1985; Hough et al. 1996; Aitken et al. 1997; Smith et al. 2000).
However, Jiang et al. (2005) used NIR adaptive optics-aided, very high-resolution imaging polarimetry 
to successfully reveal an outflow/disk system around BN. 

In this paper, we present the aperture polarimetry in the $J, H,$ and $Ks$ bands of $\sim 500$ stars of the ONC 
with the polarimetry mode of the IRSF/SIRIUS instrument (SIRPOL).
We also discuss possible intrinsically highly polarized sources and
the polarization of young brown dwarfs.

\section{Observations}
We carried out NIR polarimetric observations of M42 using 
the imaging polarimeter SIRPOL, the polarimetry mode of the SIRIUS camera
on the 1.4-m telescope IRSF (Infrared Survey Facility), 
at the South African Astronomical Observatory (SAAO), 
on the night of December 26, 2005.

SIRPOL consists of a single-beam polarimeter and the $JHKs$ simultaneous imaging camera 
SIRIUS (Nagayama et al. 2003), 
which has three 1024 x 1024 HgCdTe infrared detectors (HAWAII array).
SIRPOL enables deep, wide-field (7$\farcm$7 $\times$ 7$\farcm$7 
with a scale of 0$\farcs$45 pixel$^{-1}$), and simultaneous imaging polarimetry at $JHKs$. 
The polarimeter unit consists of a half-wave plate rotator and a polarizer installed on the upper 
side of the SIRIUS camera at the Cassegrain focus.
This minimizes instrumental polarization due to the inclined mirrors inside of the camera.
The polarization efficiencies are 95.5\%, 96.3\% and 98.5\% at J, H, and Ks respectively,
which have been used to correct the observed degrees of polarizations.
Note that the instrumental polarization is less than 0.3\%, which has been measured during SIRPOL commising run, and does not depend on the field-of-view and wavelength. 
See Kandori et al. (2006) for more details regarding the polarimeter.

The total integration time per wave plate position was 900 s.
The exposures were performed at four position angles (P.A.s) of the half-wave plate, 
with a sequence of P.A. = 0$^\circ $, 45$^\circ $, 22.5$^\circ $, and 67.5$^\circ $ 
to measure the Stokes parameters.
The seeing size was $\sim $1$\farcs$5 at J, H, and Ks.

The polarimetric results on the nebulous components and some of the point-like sources
in M42 are presented in Tamura et al. (2006a).
In this paper, we discuss $JHKs$ bands point-like sources for aperture polarimetry,
with the main goal of obtaining information on magnetic field structures.

After image calibrations in the standard manner using the IRAF (dark subtraction, 
flat-fielding with twilight flats, bad-pixel substitution, and sky subtraction),
we carried out software aperture polarimetry of point-like sources 
on the combined intensity images for each wave plate angle 
($I_{0^\circ },I_{22.5^\circ },I_{45^\circ },I_{67.5^\circ }$).
This was because the center of point sources (i.e., the aperture center) 
cannot be determined satisfactorily on the $Q$ and $U$ images.

Software aperture polarimetry was performed for several point sources in the field of view.
For the source detection and photometry, we used the IRAF DAOPHOT package (Stetson 1987).
First, we subtracted the smooth nebulous component from the original $I$ image
using a median smoothed image.
Second, we detected the positions of point-like sources from the subtracted image with the DAOFIND task.
The point-like sources have a peak intensity greater than 10$\sigma$ above the local sky on the subtracted image.
The spurious sources in automatic detection were rejected by eyes.
Finally, we carried out software aperture polarimetry of detected point-like sources on the original images.
The aperture radius was 3 pixels.
Then the local background was subtracted using the mean of a circular annulus 
around the source. 

We rejected all sources whose photometric errors were greater than 0.1 mag.
We determined a photometric zero-point by using the 
2MASS Point-Source Catalog (Cutri et al. 2003) instead of observing photometric standard stars.
We compared the 2MASS-PSC with bright sources whose instrumental magnitude errors were smaller than 0.05 mag.
The colors of our photometry were not transformed into the 2MASS system;
our photometric system was
the IRSF system.

%
After the image calibrations and photometry, %
the Stokes parameters $(I,\, Q,\, U)$, and the degree of polarization $P$ 
were calculated as follows:
\[ Q = I_{0} - I_{45}, \\  U = I_{22.5} - I_{67.5}, \]
\[ I = (I_{0} + I_{45} + I_{22.5} + I_{67.5} ) / 2,\] 
\[ P_0 = \sqrt{(Q/I)^2 + (U/I)^2}\]
\[ \theta = 0.5 \times \mathrm{atan}\frac{U}{Q} .\]
Since the polarization degree is a positive quantity, 
the derived $P_0$ values tend to be overestimated, 
especially for low signal-to-noise (S/N or P/{$\delta P$}) sources. 
The polarization degrees were debiased (Wardle \& Kronberg 1974),
using the following equation:
\[ P = \sqrt{{P_0}^2 - (\delta P)^2}.\]

\begin{figure}[htbp]
 \begin{minipage}{0.5\hsize}
  \begin{center}
   \includegraphics[width=80mm]{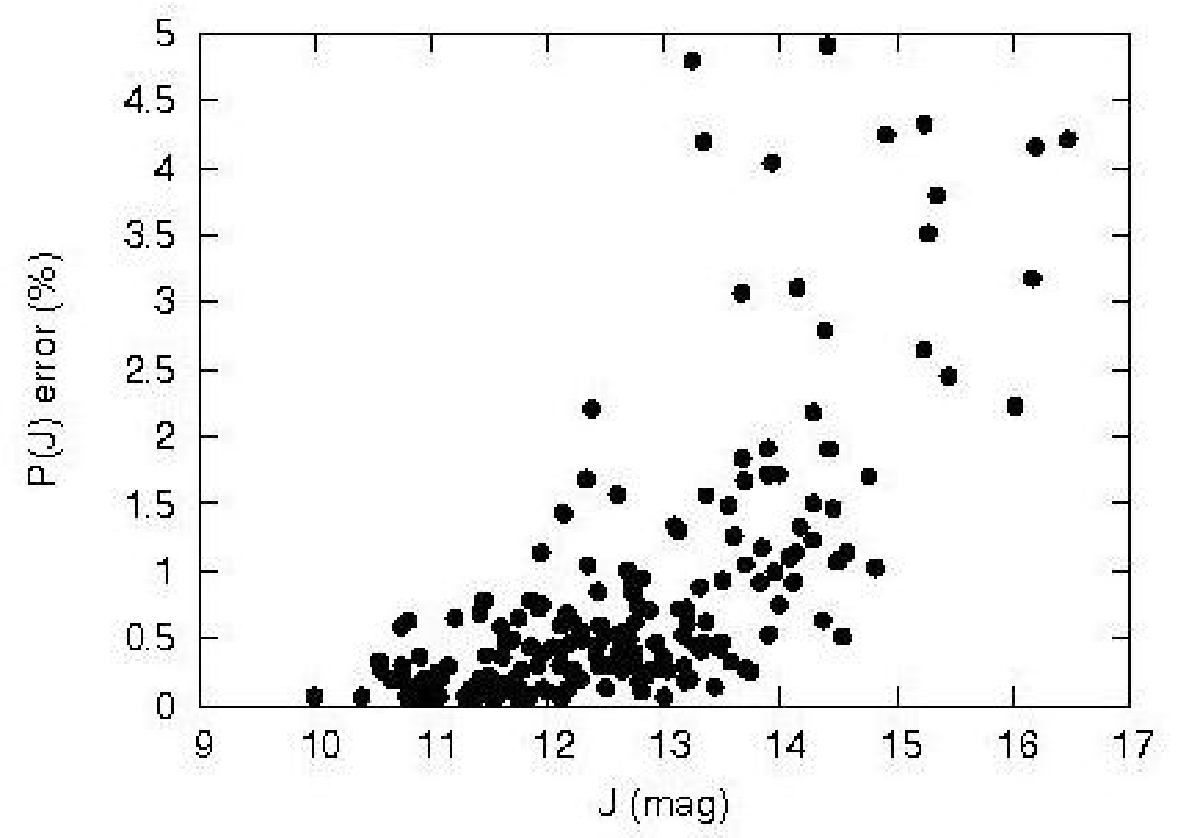}
  \end{center}
 \end{minipage}
 \begin{minipage}{0.5\hsize}
 \begin{center}
  \includegraphics[width=80mm]{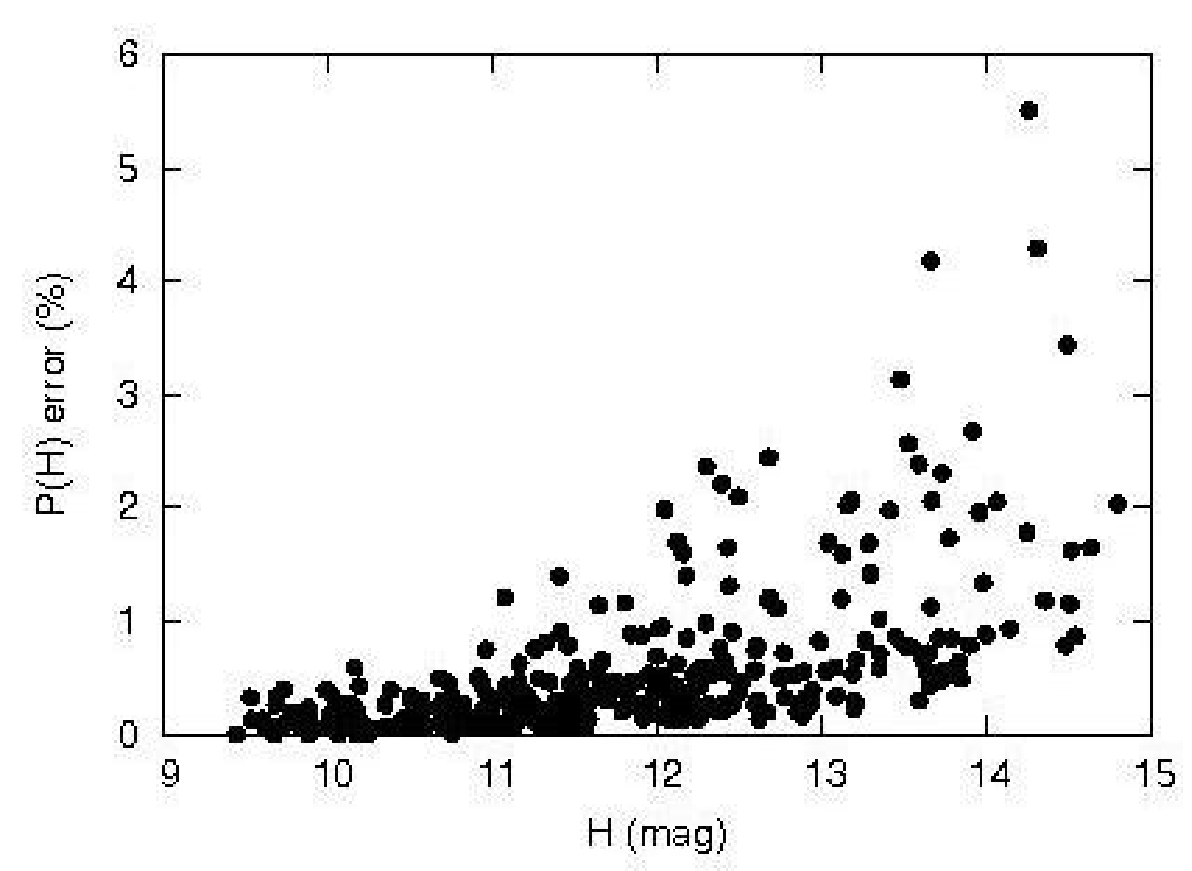}
 \end{center}
 \end{minipage}
 \begin{minipage}{0.5\hsize}
 \begin{center}
  \includegraphics[width=80mm]{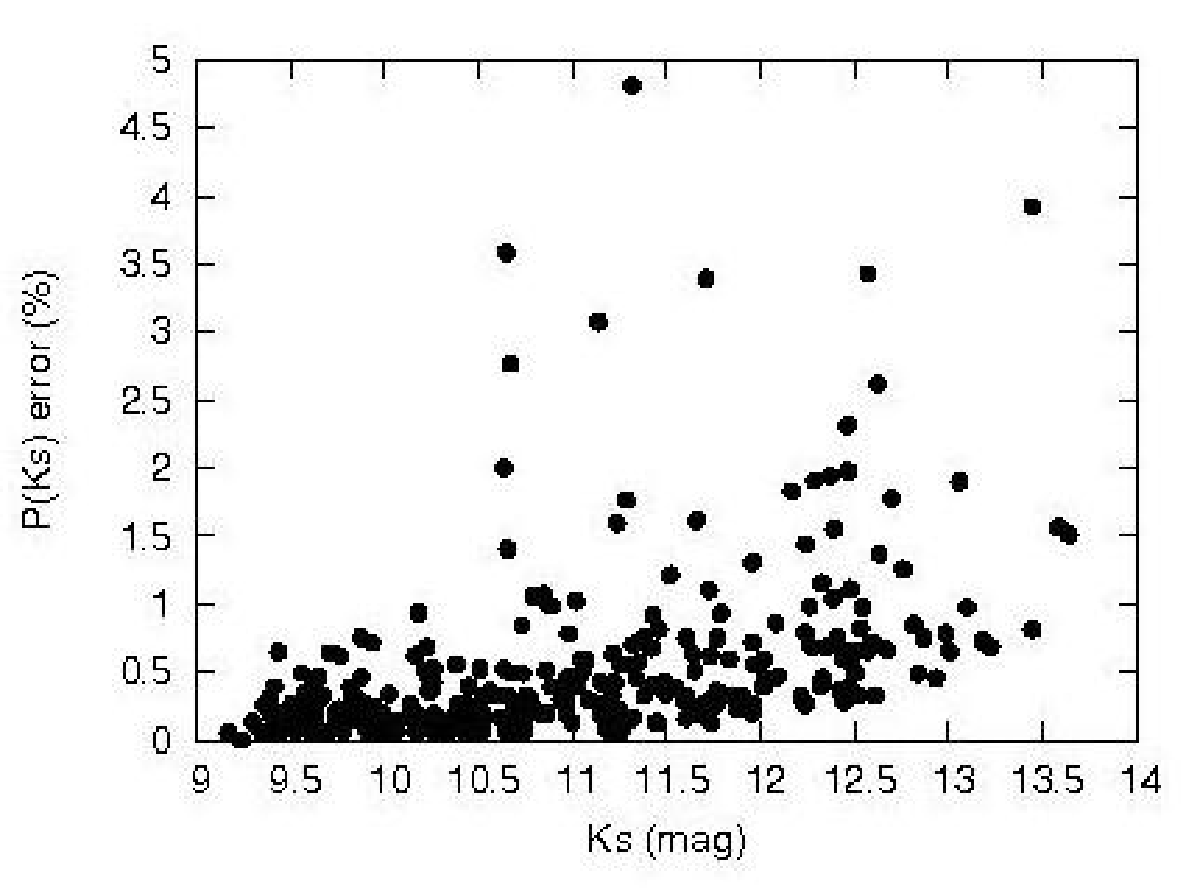}
 \end{center}
 \end{minipage}
 \caption{JHKs magnitude vs. polarization errors. 200, 314 and 279 sources 
having polarization signal-to-noize larger than 2 are plotted 
on $J$, $H$ and $Ks$ bands, respectively.}
\label{jhkerr}
\end{figure}

In total, we measured the polarizations of 
417, 498, and 483 sources, of which 200, 314, and 279 
sources have a polarization S/N larger than 2 
($P/{\delta P} > 2$),  at $J$, $H$, and $Ks$, respectively.
Figure \ref{jhkerr} shows the distribution of 
magnitude versus polarization errors of sources having a polarization S/N larger than 2.
Although several sources have large errors due to
high local sky background, there is a trend of increasing
erros toward fainter sources. 
The list of 314 sources in the $H$ band is presented in Table 1.
%

\section{Results and Discussion}

\subsection{NIR Aperture Polarization -- General Features}
Aperture polarimetry of stars provides important information 
about the direction of the magnetic fields.
If we assume the normal grain alignment mechanism, that is, that the spin axis of elongated dust grains aligns parallel to the magnetic field (Davis \& Greenstein 1951), the direction of the magnetic fields projected onto the sky can be inferred 
from the direction of the polarization vectors of stars 
($B$ $\parallel $ $E$, e.g., Weintraub, Goodman, \& Akeson 2000).
It has been proposed that other mechanisms such as the relative gas-grain
motion (Gold 1952; Lazarian 1997) or radiation (Onaka 2000; Cho et al. 2003) 
might be operating in the Orion region, but here we assume that paramagnetic
relaxation for thermally or suprathermally rotating grains
is responsible for the grain alignment (see Purcell 1979).

\begin{figure}[htb]
\includegraphics[width=14cm]{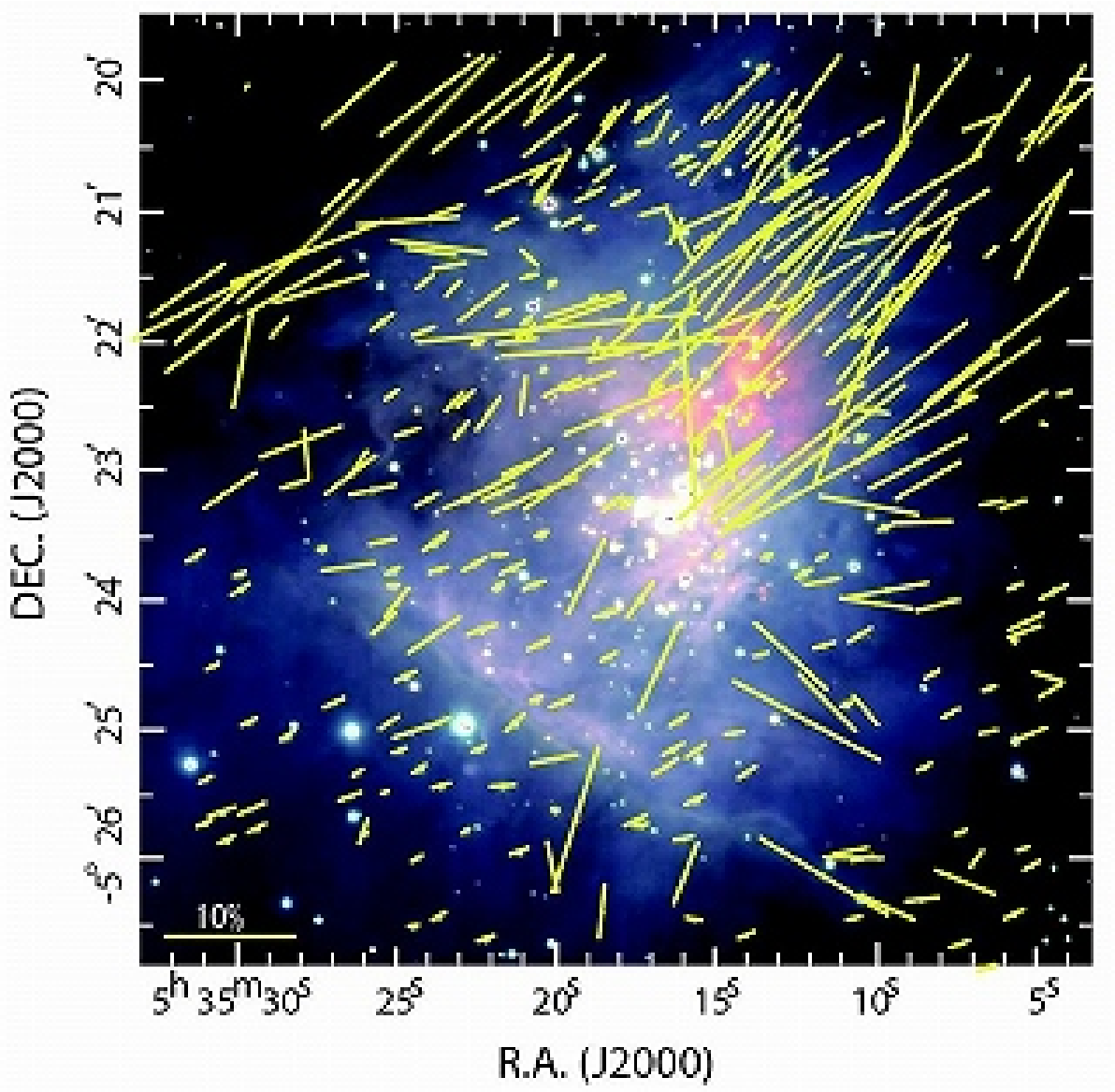}
\caption{$H$ band polarization vectors for 314 sources,
overlaid on a $JHKs$ composite color image.
10\% vector is shown at the bottom-left.}
\label{nir}
\end{figure}

In Figure \ref{nir}, we show the distribution of 314 polarization vectors for 
which we measured the aperture polarizations in the $H$ band.
We discuss mainly $H$ band data for the aperture polarimetry, 
as the extended nebula contamination is less than in the $J$ band and 
the dichroic polarization 
is higher than in the $Ks$ band. 
We confirmed that 99\% and $\sim$85\% of the sources detected at 
either $J$ or $K$ show the same polarization angle at $H$
with $\Delta$$\theta$ $<$ 30$^\circ$ and $\Delta$$\theta$ $<$ 10$^\circ$, respectively.

We note that 173 sources whose polarization S/N are larger than 2 were detected in all of the $J$, $H$, and $Ks$ bands, which can be used to check the wavelength dependence of polarization. 
The linearly fitted slopes of $P(H)$ vs. $P(J)$ and $P(Ks)$ vs. $P(H)$ diagrams are 0.65 and 0.60, with the correlation coefficients of 0.96 and 0.96, respectively. 
These slopes are consistent with the empirical values of 0.62 and 0.61 from the relation 
$P(\lambda ) \propto \lambda ^{-\alpha}$, where ${\alpha = 1.8 \pm 0.2}$ (Whittet 1992). 
On the other hand, the slopes of the only highly polarized sources (31 HPSs; see \S 3.4) are 0.64 and 0.62, with the correlation coefficients of 0.86 and 0.91, respectively. 
Thus, there is no clear difference between the slope of all sources and that of HPSs.
%

%

Although a small number of locally deviated vectors are evident, 
the impression is that the polarization vectors are relatively well aligned
in the northwest and southeast directions
in a large scale, with a gradual change over the field of view.
The average value of the NIR polarization angle is $\sim 120^\circ$.
This direction is in excellent agreement with
the ``mean field'' of the magnetic fields pervading the large-scale Orion A cloud
as revealed by submillimeter and far-infrared polarization mapping
(Schleuning 1998; Houde et al. 2004).
Even this simple comparison can lead us to conclude that
most of the NIR polarization vectors are dominated by
the dichroic polarization due to magnetically aligned dust grains,
thus tracing the magnetic fields 
between the sources and the cloud surface
within the Orion molecular cloud.

Near the OMC-1 region, we can also see an hourglass structure of polarization vectors
centered around the BN/KL region, 
which has been alleged to result from gravitational distortion 
caused by the IRc2.
These are discussed in more detail in section 3.3.

Note that the stars with lower degrees of polarization
are distributed all over the field, while the relatively highly
polarized stars are seen only around the northwest region, including
the OMC-1, and the northeast region (Figures 2 and 4).
In particular, the polarization degrees are very low to the southeast of the Orion bar and near the western edge of the frame, which might trace the magnetic field near the cloud 
surface where the extinction is relatively low.

\subsection{Magnetic Field Structures: NIR vs. FIR and Submillimeter}

Houde et al. (2004) studied the magnetic field structure of Orion A based on 
the 350 $\mu$m dust continuum observations using the Hertz polarimeter 
and SHARC II (Submillimeter High Angular Resolution Camera II)
on the CSO 10.4-m telescope.
Among several far-infrared and submillimeter polarimetric observations of this region,
this is the most extensive ($\sim $9$'$ $\times $6$'$) and 
highest resolution (12$''$ and 20$''$) data set thus far.
In contrast, our NIR data are sampled discretely
with a pencil beam
toward each star in the $\sim$8$'$ $\times$ 8$'$ region.
 
We overlaid their polarization map (rotated by 90$^\circ$) in Figure \ref{nir_sm},
which shows the 350 $\mu$m data (red lines) and $H$ band vectors (blue lines) 
on the 350 $\mu$m contour map (D. C. Lis et al. 1998).
The overall correlation between the NIR and submillimeter polarizations
is very high, not only on an average, but also on a position-by-position basis.
The best correlation is seen toward OMC-1 and its surrounding
region (including the hourglass-shaped geometry mentioned above),
toward the southwest region with a higher polarization and near 
the Orion bar with a relatively low polarization.

Schleuning et al. (1998) reported that the highest polarization at
350 and 100 $\mu$m was well away from IRc2 to the northwest,
in a region not associated with any flux features.
This high polarization region seen at far-IR and submillimeter wavelengths
is also coincident with the NIR high-polarization region,
suggesting that the magnetic field lines are running
almost perpendicular to the line-of-sight in this region;
therefore, the magnetic field projected on the sky
is seen as a maximum here.

The hourglass-shaped magnetic field pattern centered near IRc2
is expected from theoretical models in which the magnetic field
lines are dragged along with the contracting gas and then
wound up by rotation in a disk (Galli \& Shu 1993).
Our NIR data, as has been suggested by previous far-IR and submillimeter
polarimetry, also confirm that such a magnetic field geometry
applies to the region of a radius of $\sim$0.5 pc (240$''$) around IRc2.

However, note also that several clear differences exist:
one is the northeast region where both polarizations are
relatively higher but their position angles are systematically different
by $\sim$40$^\circ$ (NIR angles are larger).
Another difference is seen just to the southeast of the Orion bar,
where the smaller NIR polarizations are almost perpendicular
to the larger NIR or FIR/submm polarizations.
It appears that several field components are seen in these regions.

In Figure \ref{MG91} we show histograms of the polarization position angles of stars in M42. 
Each wavelength datum, including the submillimeter data,
is well fitted with the modeling employed by Myers \& Goodman et al. (1991).
The mean direction of the NIR polarization vectors
almost exactly coincides with the submillimeter direction,
confirming the impression described in Section 3.1.

To verify the apparent similarity of the distributions shown in Figure \ref{MG91}, 
the observed distributions of polarization position angles can be modeled 
as the sum of the contributions from a uniform and nonuniform magnetic field 
using the technique described by Myers \& Goodman (1991)(hereafter MG91), 
which is similar to the method first proposed by Chandrasekhar \& Fermi (1953).
The results of this fitting are summarized in Table \ref{PA}.

The parameters $< \theta >$ and $s$ in Table \ref{PA} are the mean and dispersion of the distribution, 
respectively, based on least-squares fits of the MG91 model ($f^{3D}$ in MG91).
The average position angles are identical among $JHKs$ and 350 $\mu$m.
Therefore, we conclude that the distributions
of the NIR and submillimeter polarization are very similar to each other.

The large-scale ($\sim$30 pc) magnetic field structures
of the Orion region and the Orion A cloud are inferred from the optical polarimetry
by Breger (1976) and Vrba et al. (1988), respectively.
Their average direction is 100$^\circ $ and 110$^\circ $,
roughly consistent or slightly smaller than the above NIR and submillimeter data.
Figure \ref{2p} compares the position angle distributions of the small polarization
($P(H)$ $<$ 2\%) sources and the large polarization ($P(H)$ $>$ 2\%) sources.
Although their distributions are more or less similar to each other,
their peak position angle is slightly different by $\sim$10$^{\circ}$;
the small NIR polarization sources show a polarization angle
between the above optical and the large NIR polarization sources.
The optical polarizations trace the magnetic fields in the periphery
of the Orion cloud, while the small and large NIR polarizations
trace the fields near and inside of the cloud surface, respectively. 
Therefore, the mean magnetic field pervades the Orion region
from the giant molecular cloud scale to the cloud core scale
with only a slight change in position angles.

%
\subsection{Separating Possible Intrinsic Polarized Sources}
Note that not all of the observed NIR polarizations are caused by interstellar polarization.
Highly polarized sources occur that cannot be explained by interstellar polarization.
The highly polarized sources might possess intrinsic polarizations due to their circumstellar structures that 
are too small to be resolved with our 1.5$''$ resolution images.
Resolved circumstellar structures of M42 at several scales have been reported by Tamura et al. (2006a).
Alternatively, deep embedded sources might have large interstellar polarizations 
(Jones 1989; Jones et al. 1992).
Therefore, we cannot determine which sources are dichroic or intrinsic solely
on the basis of the degree or angle of polarization.
As a way to separate these two kinds of polarizations, 
 we employed a $P(H)/(H-K)$ value of BN as an upper limit for the interstellar polarization. 
The high degree of polarization of BN has been used to deduce that BN is polarized by dichroic absorption (e.g., Smith et al. 2000, Simpson et al. 2006).


Figure \ref{Pth_H-K} shows the degrees of polarization in the $H$ band vs. the $H-Ks$ 
color as well as the polarization angle in the $H$ band vs. the $H-Ks$ color. 
 The solid line denotes the assumed upper limit of BN, $P(H)/H-K = 8.2$.
BN is one of the most polarized stars. 
Minchin et al. (1991) measured $P(H)=31\%$ with $\theta \sim 114$ degrees in a $6 ''$ aperture.
Although Jiang et al. (2005) detected an outflow/disk system around BN from very high-resolution ($\sim$ 1000 AU scale) imaging polarimetry, its contribution to the observed IR polarization is small.
In fact, its dichroic nature has been confirmed by the wavelength shift of the polarization peak of absorption features (Kobayashi et al. 1980, Aitken et al. 1989).
Furthermore, we checked the $P(H)/E(H-K)$ and $P(H)/A(H)$ values of background field stars in other star forming regions of the Taurus cloud and the Ophiuchus cloud (Whittet et al. 2008); 
even the largest values are 4.6 in $P(H)/E(H-K)$ and 1.6 in $P(H)/A(H)$. 
Therefore our upper limit, $P(H)/E(H-K)=8.2$ or $P(H)/A(H)=2.9$, is sufficiently large to select the highly polarized sources. 
Note also that the linear fitted slope (average slope) of the $P(H)$ vs. the $H-Ks$ diagram is $P(H)/ (H - Ks) = 2.7$. 
The slope value of BN is three-times larger than the average value in the M42 region.

In fact, as expected, most of the bluer ($H-Ks \leq 0.5$) sources have $P < P_{\mathrm{max}}$, 
which confirms the intracloud origin of the NIR polarizations.
This also excludes other possible sources of polarization
such as gas motion or radiation because they cannot explain
the color dependence on polarization.
Furthermore, we observe no clear turnover of $P(H)$ even up to $H-Ks$ $\sim$2.5 ($A_{V}$ $\sim$36 mag).
This also excludes a weak alignment of dust grains in the dense cloud
as suggested by Goodman et al. (1995), and
supports the notion that most, if not all, of the observed NIR polarizations
indeed trace magnetic fields in the Orion molecular cloud. 
However, several sources show polarization larger than the ``upper limit''.

Sources with larger $P(H)/(H-Ks)$ values are better candidates for
having intrinsic polarizations.
These sources are discussed further in the next section.
We therefore reject 51 sources that are likely to have intrinsic polarizations. 
Note that the rejected sources do not affect the previous discussion
on the magnetic fields in this region.

To verify the apparent similarity of the distributions shown 
in Figure \ref{nir_sm}, we subtracted the mean angle of the submillimeter data 
in the circumference less than 20$''$ from each NIR polarization angle.
Figure \ref{lowP} shows the histogram of the subtracted angles.
This histogram clearly denotes a peak at 0$^{\circ}$.
If we assume that the 350 $\mu$m polarization vectors indicate the magnetic field, 
after the subtraction of possible intrinsic polarized sources,
the dichroic polarization at NIR wavelengths also conforms well to the magnetic
field structures inside these dense molecular clouds.

\subsection{Highly Polarized Sources}

We found 51 sources having a lager NIR polarization (excess polarization) than interstellar polarization.
The spatial distribution of the 51 highly polarized sources (HPSs) selected in the previous section is shown in Figure \ref{Pmax_c}.
Note that 31 HPSs were detected in all of the $J$, $H$ and $Ks$ bands.
These sources are mostly distributed in the northwest, including near the OMC-1 region.
The outflows produced by both massive and low-mass YSOs are or were powerful enough to open a cavity within their parent cloud core, in a direction that tends to be perpendicular to the dense, disklike structures around the star (Tamura et al. 2006a).

The high polarization we detected could most likely to arise from scattered light in the local circumstellar environment (such as a cavity evacuated by the outflow from YSOs).
Additional high-resolution ($<$1$''$) studies at various wavelengths of these sources are necessary.

%

Our method to detect HPSs (based on $P$ vs. $A(V)$) tend to select sources with their intrinsic polarization angle parallel to the direction of magnetic field; 
this is because the sources whose intrinsic polarization perpendicular to this cause a depolarization effect.
For this reason, the number of HPSs (51 sources) detected in our data serves as lower limit.

We found that selected 51 HPSs are characterized by the following features:
(1) The polarization angle of HPSs is roughly aligned with magnetic field.
(2) The excess polarization of HPSs with respect to the employed upper limit of interstellar
polarization is not large as shown in Fig. \ref{Pth_H-K}. 
These features indicate that the intrinsic polarization associated with HPSs is not large, 
comparable or less than the interstellar polarization, so that they show moderate excess 
polarization with the angle roughly aligned with magnetic field.
In order to evaluate the deviation of HPS's polarization angles from magnetic fields, 
we carried out simple calculations of the superposition of two polarization vectors, 
intrinsic polarization (P(INT), PA(INT)) and interstellar polarization (P(ISM), PA(ISM)). 
Assuming the PA(INT) to be randomly distributed, we calculated
$\Delta$PA = rms(PA(OBS) - PA(ISM)) for the different values of P(ISM)/P(INT),
where PA(OBS) shows the polarization angle after the superposition. 
The $\Delta$PA shows variation in the direction of the observed (superposed) vector and
magnetic field. 
We found $\Delta$PA $< 14^{\circ}$ when P(ISM)/P(INT) $>$ 1.5, indicating
that the polarization angles of HPSs with moderate excess polarizations 
tend to be aligned with the direction of magnetic fields. 
We note that the deviation angle cannot be detected with our data whose accuracy in polarization
angle is $< 14.3^{\circ}$ (i.e., P/$\Delta$P $>$ 2). 
These results can naturally explain the observed polarization features of HPSs described above.

Here we stress that several low-luminosity sources, presumably young brown dwarfs,
are included in the ``highly polarized sources''.
Brown dwarfs are stars less than 0.08 solar masses that are unable to sustain 
hydrogen fusion in their interiors. 
Recent observations suggest that young brown dwarfs appear to have
circumstellar disks like their massive counterparts, T Tauri stars,
but these disks are too small or faint to be resolved (e.g., Jayawardhana et al. 2003).
Our NIR polarimetry can establish a constraint on their circumstellar structure,
even if they are not resolved (see Section 3.4).

It is theoretically expected that gravitation of low-mass stars 
(including brown dwarfs) produce a more flared (with a larger-scale height) 
disk than those of massive stars (Walker et al. 2004). 
Such sources are observed only in scattered light.
These might be observed as highly polarized sources.
The disk obscuring probability of brown dwarfs in the $K$ band (2.2$\mu$m) 
is estimated to be larger than that for Class II stars by a factor of $\sim$ 1-3 (Walker et al. 2004).
Thus, a stellar mass dependence of HPS detection rate could be expected.

It is not simple to select young brown dwarfs from our $JHKs$ Stokes $I$ parameters alone;
Therefore, we cross-correlated our highly polarized sources
with known young brown dwarf candidates and low-mass stars
whose masses have been 
estimated from both photometry and spectroscopy (Slesnick et al. 2004, 
Riddick et al. 2007; see Table \ref{AvMP_S},\ref{AvMP_R}).

In Figure \ref{mass}, we compare the polarization efficiency, $P(H)/A(H)$, 
with each stellar/substellar mass.
The horizontal line indicates the assumed upper limit of the ``interstellar'' polarization calculated with $P(H) = 8.2 (H-Ks)$.
At a first glance, we cannot see any clear correlation in this figure.
However, note that almost all of the young brown dwarf candidates (nine sources) 
whose masses are less than 0.08 $M_{\odot}$ 
and polarizations are positively measured are plotted over
the upper limit line; thus, their polarization is more likely to be 
due to scattering rather than to their being dichroic.
We consider this to be evidence for the presence of compact
circumstellar structures such as disk/outflow systems around these young brown dwarf candidates
(see also, Tamura et al. 2006a and 2006b; Kandori et al. 2007).

\section{Conclusion}
We conducted deep $JHKs$ imaging polarimetry of a 7$\farcm$7 $\times$ 7$\farcm$7 area 
of M42 and here present the results of the aperture polarimetry. 
Our main conclusions are summarized as follows:

\begin{enumerate}

\item Most of the NIR polarizations are dichroic.
Their global and local vector patterns are in good
agreement with previous FIR and submillimeter polarization patterns,
except for the 90$^{\circ}$ difference of position angles.

\item A positive correlation exists between $P(H)$ and $H-Ks$;
most of $P(H)$ is below $P_{max}$, except for possible
highly polarized sources.
This supports an intracloud origin of the NIR polarizations,
excluding (a) a non-intracloud magnetic origin of polarization,
and (b) a weaker alignment of dust grains in the cloud.

\item We argue that both NIR dichroic polarizations
and FIR/submillimeter thermal polarizations trace the magnetic fields
in the OMC-1 region.

\item The magnetic fields are pinched at scales less than 0.5 pc 
with a centroid near IRc2.
The hourglass-shaped magnetic field pattern
is explained by models in which the magnetic field
lines are dragged along with the contracting gas and then
wound up by rotation in a disk.

\item The highly polarized region to the northwest of IRc2
and 
the low-polarized region near the bright bar,
commonly seen in both NIR and FIR/submillimeter data,
are explained by the field geometry;
the latter is the field along and the former perpendicular to
the line of sight.

\item We also discriminated $\sim$50 possible highly polarized sources
whose polarizations are more likely to be intrinsic
rather than dichroic.
Their polarization efficiencies ($P(H)/A(H)$) are too large
to be explained by normal interstellar polarization.

\item For 9 young brown dwarf candidates,
we also suggest the existence of higher polarization efficiency,
which may present geometrical evidence for (unresolved) circumstellar structures
around young brown dwarfs.

\end{enumerate}

We are grateful to Shuji Sato for helpful suggestions.
Thanks are due to the staff in SAAO for their kind help during the observations.
We thank D. C. Lisfor kindly providing the 350 ƒÊm map in FITS format. 
We also thank Noboru Ebizuka Tetsuo Nishino, and Toshihide Kawai 
for their technical support in the development of SIRPOL.
The IRSF/SIRIUS project was initiated and supported by Nagoya
University, National Astronomical Observatory of Japan,
and The University of Tokyo in collaboration with South
African Astronomical Observatory under a financial support
of Grants-in-Aid for Scientific Research on Priority Area
(A) No. 10147207 and No. 10147214, and Grants-in-Aid
No. 13573001 and No. 16340061 of the Ministry of Education,
Culture, Sports, Science, and Technology of Japan.
This study was partly supported by a MEXT Grant-in-Aid for
Scientific Research in Priority Areas, gDevelopment of Extrasolar
Planetary Science,h and by grants-in-aid fromMEXT(Nos.
16077204, 16077171, and 19204018).

\clearpage
\begin{center}
\begin{deluxetable}{llcccc}
\tablecolumns{9}
\tablewidth{0pc}
\tablecaption{ H band polarization list of 314 sources.}
\tablehead{
\colhead{RA (J2000)} & \colhead{Dec (J2000)} & \colhead{\textit{H} (mag)} & \colhead{\textit{P} (\%)}  & \colhead{$\theta$ (deg)}  & \colhead{ID}
}
\startdata
05 35 22.20	&	-05 26 37.3	&	10.43 	&	1.3	$\pm$	0.1 	&	109 	$\pm$	1	&	1	\\
05 35 06.60	&	-05 26 50.8	&	11.50 	&	1.4	$\pm$	0.1 	&	108 	$\pm$	1	&	2	\\
05 35 24.46	&	-05 26 31.4	&	10.86 	&	0.7	$\pm$	0.1 	&	99 	$\pm$	3	&	3	\\
05 35 14.07	&	-05 26 35.8	&	12.60 	&	1.3	$\pm$	0.6 	&	113 	$\pm$	12	&	5	\\
05 35 23.66	&	-05 26 27.0	&	11.40 	&	1.2	$\pm$	0.1 	&	121 	$\pm$	2	&	7	\\
05 35 07.45	&	-05 26 40.0	&	12.25 	&	1.5	$\pm$	0.1 	&	112 	$\pm$	3	&	8	\\
05 35 15.72	&	-05 26 28.3	&	11.43 	&	1.2	$\pm$	0.1 	&	80 	$\pm$	3	&	11	\\
05 35 18.58	&	-05 26 24.8	&	12.40 	&	4.1	$\pm$	0.3 	&	174 	$\pm$	2	&	12	\\
05 35 10.74	&	-05 26 28.0	&	12.08 	&	1.1	$\pm$	0.3 	&	102 	$\pm$	8	&	14	\\
05 35 09.66	&	-05 26 23.3	&	11.33 	&	4.4	$\pm$	0.2 	&	66 	$\pm$	1	&	18	\\
05 35 05.74	&	-05 26 26.1	&	10.90 	&	1.6	$\pm$	0.1 	&	115 	$\pm$	1	&	19	\\
05 35 08.60	&	-05 26 19.4	&	12.26 	&	2.5	$\pm$	0.2 	&	97 	$\pm$	2	&	21	\\
05 35 20.13	&	-05 26 04.1	&	13.03 	&	3.9	$\pm$	0.6 	&	12 	$\pm$	4	&	22	\\
05 35 24.34	&	-05 26 00.3	&	12.38 	&	0.9	$\pm$	0.2 	&	122 	$\pm$	6	&	23	\\
05 35 15.97	&	-05 26 07.2	&	12.27 	&	5.0	$\pm$	0.6 	&	159 	$\pm$	3	&	24	\\
05 35 11.67	&	-05 26 08.6	&	11.60 	&	11.5	$\pm$	0.3 	&	59 	$\pm$	1	&	25	\\
05 35 30.27	&	-05 25 51.6	&	11.31 	&	1.2	$\pm$	0.1 	&	115 	$\pm$	2	&	26	\\
05 35 07.27	&	-05 26 11.3	&	14.00 	&	4.9	$\pm$	0.9 	&	68 	$\pm$	5	&	27	\\
05 35 21.16	&	-05 25 56.9	&	10.51 	&	1.5	$\pm$	0.1 	&	111 	$\pm$	1	&	28	\\
05 35 29.34	&	-05 25 46.2	&	10.90 	&	1.5	$\pm$	0.1 	&	118 	$\pm$	1	&	31	\\
05 35 30.96	&	-05 25 43.9	&	11.56 	&	1.3	$\pm$	0.1 	&	115 	$\pm$	2	&	32	\\
05 35 26.00	&	-05 25 47.8	&	10.35 	&	1.9	$\pm$	0.1 	&	160 	$\pm$	1	&	33	\\
05 35 10.48	&	-05 26 00.3	&	12.24 	&	1.6	$\pm$	0.3 	&	105 	$\pm$	5	&	34	\\
05 35 10.76	&	-05 26 00.0	&	13.52 	&	4.5	$\pm$	0.8 	&	89 	$\pm$	5	&	35	\\
05 35 05.19	&	-05 26 03.6	&	12.85 	&	1.4	$\pm$	0.2 	&	118 	$\pm$	4	&	36	\\
05 35 22.83	&	-05 25 47.6	&	11.91 	&	0.7	$\pm$	0.2 	&	122 	$\pm$	10	&	37	\\
05 35 06.91	&	-05 26 00.5	&	12.67 	&	1.5	$\pm$	0.2 	&	144 	$\pm$	4	&	38	\\
05 35 10.84	&	-05 25 56.9	&	12.39 	&	2.7	$\pm$	0.2 	&	101 	$\pm$	2	&	39	\\
05 35 30.42	&	-05 25 38.6	&	10.74 	&	2.1	$\pm$	0.1 	&	110 	$\pm$	1	&	40	\\
05 35 16.04	&	-05 25 51.0	&	12.33 	&	1.4	$\pm$	0.5 	&	101 	$\pm$	10	&	41	\\
05 35 29.48	&	-05 25 37.2	&	13.20 	&	2.2	$\pm$	0.2 	&	114 	$\pm$	3	&	44	\\
05 35 05.46	&	-05 25 55.8	&	13.21 	&	0.9	$\pm$	0.3 	&	109 	$\pm$	8	&	47	\\
05 35 17.36	&	-05 25 44.7	&	10.57 	&	1.1	$\pm$	0.3 	&	101 	$\pm$	8	&	48	\\
05 35 19.39	&	-05 25 42.3	&	12.61 	&	11.7	$\pm$	0.8 	&	163 	$\pm$	2	&	49	\\
05 35 17.55	&	-05 25 42.8	&	10.94 	&	1.8	$\pm$	0.4 	&	116 	$\pm$	7	&	51	\\
05 35 20.04	&	-05 25 37.6	&	10.52 	&	0.5	$\pm$	0.1 	&	89 	$\pm$	4	&	53	\\
05 35 26.42	&	-05 25 31.5	&	12.86 	&	1.8	$\pm$	0.3 	&	111 	$\pm$	4	&	54	\\
05 35 07.40	&	-05 25 48.1	&	12.21 	&	0.9	$\pm$	0.2 	&	112 	$\pm$	7	&	55	\\
05 35 25.34	&	-05 25 29.5	&	12.09 	&	0.6	$\pm$	0.2 	&	108 	$\pm$	10	&	59	\\
05 35 30.92	&	-05 25 23.6	&	12.14 	&	1.1	$\pm$	0.1 	&	118 	$\pm$	4	&	62	\\
05 35 23.59	&	-05 25 26.4	&	11.44 	&	1.4	$\pm$	0.2 	&	106 	$\pm$	4	&	64	\\
05 35 15.68	&	-05 25 33.1	&	10.51 	&	1.6	$\pm$	0.3 	&	116 	$\pm$	5	&	65	\\
05 35 21.66	&	-05 25 26.5	&	10.83 	&	1.8	$\pm$	0.1 	&	94 	$\pm$	2	&	66	\\
05 35 26.19	&	-05 25 20.4	&	12.31 	&	0.9	$\pm$	0.3 	&	123 	$\pm$	9	&	67	\\
05 35 24.25	&	-05 25 18.7	&	10.42 	&	1.1	$\pm$	0.1 	&	126 	$\pm$	2	&	69	\\
05 35 14.06	&	-05 25 20.4	&	11.67 	&	1.6	$\pm$	0.7 	&	124 	$\pm$	12	&	73	\\
05 35 16.58	&	-05 25 17.7	&	11.96 	&	2.2	$\pm$	0.5 	&	123 	$\pm$	7	&	74	\\
05 35 24.89	&	-05 25 10.2	&	11.89 	&	1.1	$\pm$	0.3 	&	112 	$\pm$	7	&	75	\\
05 35 20.05	&	-05 25 14.3	&	12.16 	&	3.7	$\pm$	1.6 	&	101 	$\pm$	13	&	76	\\
05 35 22.41	&	-05 25 09.5	&	11.24 	&	1.1	$\pm$	0.3 	&	114 	$\pm$	7	&	79	\\
05 35 28.38	&	-05 25 03.4	&	10.51 	&	1.1	$\pm$	0.1 	&	124 	$\pm$	2	&	81	\\
05 35 08.53	&	-05 25 17.9	&	11.13 	&	1.1	$\pm$	0.1 	&	119 	$\pm$	2	&	83	\\
05 35 15.66	&	-05 25 10.5	&	11.74 	&	1.8	$\pm$	0.4 	&	125 	$\pm$	6	&	84	\\
05 35 16.36	&	-05 25 09.7	&	10.99 	&	1.9	$\pm$	0.2 	&	120 	$\pm$	3	&	85	\\
05 35 24.51	&	-05 25 01.5	&	12.17 	&	2.9	$\pm$	0.3 	&	109 	$\pm$	3	&	86	\\
05 35 29.58	&	-05 24 56.8	&	10.99 	&	1.3	$\pm$	0.1 	&	115 	$\pm$	1	&	88	\\
05 35 25.21	&	-05 24 57.2	&	13.46 	&	1.7	$\pm$	0.9 	&	125 	$\pm$	14	&	91	\\
05 35 23.67	&	-05 24 57.3	&	11.47 	&	1.8	$\pm$	0.3 	&	114 	$\pm$	5	&	92	\\
05 35 21.29	&	-05 24 57.3	&	10.02 	&	1.9	$\pm$	0.3 	&	133 	$\pm$	5	&	93	\\
05 35 14.46	&	-05 25 02.1	&	11.35 	&	1.4	$\pm$	0.5 	&	98 	$\pm$	9	&	94	\\
05 35 26.93	&	-05 24 48.1	&	12.51 	&	1.7	$\pm$	0.5 	&	115 	$\pm$	8	&	96	\\
05 35 16.20	&	-05 24 56.3	&	10.95 	&	1.6	$\pm$	0.2 	&	122 	$\pm$	4	&	97	\\
05 35 05.29	&	-05 25 02.9	&	13.35 	&	2.8	$\pm$	0.6 	&	112 	$\pm$	6	&	100	\\
05 35 06.54	&	-05 25 01.5	&	12.91 	&	1.3	$\pm$	0.3 	&	103 	$\pm$	7	&	101	\\
05 35 12.20	&	-05 24 56.4	&	12.46 	&	12.9	$\pm$	0.9 	&	62 	$\pm$	2	&	102	\\
05 35 19.84	&	-05 24 47.9	&	11.48 	&	1.9	$\pm$	0.5 	&	113 	$\pm$	7	&	104	\\
05 35 20.63	&	-05 24 46.4	&	10.15 	&	0.9	$\pm$	0.3 	&	112 	$\pm$	8	&	105	\\
05 35 10.37	&	-05 24 51.4	&	13.98 	&	2.9	$\pm$	1.3 	&	40 	$\pm$	13	&	108	\\
05 35 07.71	&	-05 24 52.8	&	13.08 	&	1.3	$\pm$	0.6 	&	119 	$\pm$	13	&	110	\\
05 35 10.94	&	-05 24 48.7	&	10.23 	&	1.0	$\pm$	0.1 	&	115 	$\pm$	2	&	112	\\
05 35 30.71	&	-05 24 31.2	&	12.08 	&	1.1	$\pm$	0.2 	&	117 	$\pm$	5	&	113	\\
05 35 11.56	&	-05 24 48.1	&	13.12 	&	6.2	$\pm$	1.2 	&	45 	$\pm$	6	&	114	\\
05 35 24.69	&	-05 24 35.7	&	10.46 	&	1.7	$\pm$	0.1 	&	115 	$\pm$	2	&	115	\\
05 35 06.52	&	-05 24 41.4	&	10.60 	&	1.0	$\pm$	0.1 	&	110 	$\pm$	2	&	120	\\
05 35 16.81	&	-05 24 32.1	&	12.39 	&	7.3	$\pm$	2.2 	&	155 	$\pm$	9	&	121	\\
05 35 04.61	&	-05 24 42.4	&	11.57 	&	2.0	$\pm$	0.1 	&	122 	$\pm$	1	&	122	\\
05 35 18.37	&	-05 24 26.8	&	11.64 	&	1.4	$\pm$	0.6 	&	118 	$\pm$	12	&	128	\\
05 35 08.30	&	-05 24 34.9	&	11.96 	&	1.3	$\pm$	0.3 	&	119 	$\pm$	6	&	129	\\
05 35 04.69	&	-05 24 35.9	&	13.73 	&	2.3	$\pm$	0.5 	&	73 	$\pm$	6	&	130	\\
05 35 20.54	&	-05 24 20.9	&	11.71 	&	1.8	$\pm$	0.5 	&	117 	$\pm$	7	&	133	\\
05 35 23.81	&	-05 24 16.8	&	12.30 	&	4.8	$\pm$	2.4 	&	125 	$\pm$	14	&	135	\\
05 35 13.61	&	-05 24 25.7	&	11.48 	&	1.0	$\pm$	0.4 	&	106 	$\pm$	12	&	136	\\
05 35 12.70	&	-05 24 26.4	&	13.13 	&	7.5	$\pm$	1.6 	&	47 	$\pm$	6	&	137	\\
05 35 22.32	&	-05 24 14.2	&	10.04 	&	2.3	$\pm$	0.2 	&	110 	$\pm$	2	&	139	\\
05 35 25.37	&	-05 24 11.4	&	12.01 	&	2.4	$\pm$	0.9 	&	134 	$\pm$	11	&	140	\\
05 35 11.65	&	-05 24 21.5	&	11.63 	&	1.7	$\pm$	0.5 	&	107 	$\pm$	9	&	143	\\
05 35 10.54	&	-05 24 16.7	&	12.38 	&	2.0	$\pm$	0.8 	&	122 	$\pm$	11	&	146	\\
05 35 05.73	&	-05 24 18.5	&	10.06 	&	0.6 $\pm$	0.1 	&	4 	$\pm$	1	&	149	\\
05 35 24.96	&	-05 24 01.5	&	11.65 	&	2.8	$\pm$	1.1 	&	141 	$\pm$	12	&	150	\\
05 35 18.47	&	-05 24 07.0	&	10.75 	&	1.8	$\pm$	0.3 	&	104 	$\pm$	5	&	151	\\
05 35 29.85	&	-05 23 55.5	&	10.85 	&	1.7	$\pm$	0.1 	&	122 	$\pm$	1	&	153	\\
05 35 05.41	&	-05 24 15.1	&	11.81 	&	2.7	$\pm$	0.3 	&	110 	$\pm$	3	&	156	\\
05 35 25.08	&	-05 23 53.8	&	10.97 	&	0.9	$\pm$	0.4 	&	95 	$\pm$	12	&	158	\\
05 35 17.38	&	-05 24 00.3	&	10.96 	&	3.4	$\pm$	0.7 	&	143 	$\pm$	6	&	159	\\
05 35 05.37	&	-05 24 10.5	&	10.70 	&	3.0	$\pm$	0.1 	&	116 	$\pm$	1	&	160	\\
05 35 29.86	&	-05 23 48.4	&	11.24 	&	1.0	$\pm$	0.1 	&	109 	$\pm$	4	&	161	\\
05 35 21.80	&	-05 23 53.9	&	9.76 	&	0.9	$\pm$	0.1 	&	123 	$\pm$	4	&	164	\\
05 35 20.67	&	-05 23 53.2	&	11.00 	&	1.6	$\pm$	0.3 	&	111 	$\pm$	6	&	166	\\
05 35 10.03	&	-05 24 01.9	&	13.29 	&	5.6	$\pm$	1.7 	&	82 	$\pm$	9	&	167	\\
05 35 08.23	&	-05 24 03.2	&	12.19 	&	2.7	$\pm$	0.3 	&	105 	$\pm$	3	&	168	\\
05 35 25.08	&	-05 23 46.8	&	10.08 	&	1.2	$\pm$	0.2 	&	118 	$\pm$	5	&	169	\\
05 35 26.49	&	-05 23 45.0	&	12.44 	&	4.1	$\pm$	1.3 	&	98 	$\pm$	9	&	170	\\
05 35 31.24	&	-05 23 40.0	&	10.78 	&	1.5	$\pm$	0.1 	&	123 	$\pm$	1	&	172	\\
05 35 07.64	&	-05 24 00.7	&	12.08 	&	0.8	$\pm$	0.1 	&	95 	$\pm$	5	&	173	\\
05 35 23.65	&	-05 23 46.3	&	11.28 	&	1.5	$\pm$	0.5 	&	108 	$\pm$	9	&	174	\\
05 35 19.07	&	-05 23 49.6	&	11.26 	&	6.3	$\pm$	0.7 	&	154 	$\pm$	3	&	176	\\
05 35 09.68	&	-05 23 55.9	&	11.65 	&	6.3	$\pm$	0.4 	&	101 	$\pm$	2	&	178	\\
05 35 21.36	&	-05 23 45.5	&	10.58 	&	2.1	$\pm$	0.3 	&	116 	$\pm$	4	&	179	\\
05 35 22.55	&	-05 23 43.7	&	10.57 	&	2.1	$\pm$	0.1 	&	119 	$\pm$	2	&	180	\\
05 35 11.72	&	-05 23 51.8	&	12.68 	&	2.7	$\pm$	1.2 	&	102 	$\pm$	13	&	181	\\
05 35 27.30	&	-05 23 36.6	&	10.00 	&	0.6	$\pm$	0.1 	&	100 	$\pm$	3	&	182	\\
05 35 04.50	&	-05 23 56.5	&	10.29 	&	2.4	$\pm$	0.1 	&	118 	$\pm$	1	&	183	\\
05 35 05.71	&	-05 23 54.1	&	11.41 	&	1.1	$\pm$	0.1 	&	114 	$\pm$	2	&	186	\\
05 35 12.28	&	-05 23 48.0	&	9.84 	&	1.3	$\pm$	0.1 	&	118 	$\pm$	3	&	188	\\
05 35 21.77	&	-05 23 39.3	&	11.38 	&	1.2	$\pm$	0.3 	&	94 	$\pm$	6	&	189	\\
05 35 25.43	&	-05 23 33.3	&	11.84 	&	2.0	$\pm$	0.5 	&	103 	$\pm$	7	&	191	\\
05 35 23.81	&	-05 23 34.3	&	10.18 	&	2.0	$\pm$	0.2 	&	121 	$\pm$	3	&	192	\\
05 35 23.65	&	-05 23 32.0	&	10.70 	&	1.6	$\pm$	0.3 	&	128 	$\pm$	5	&	193	\\
05 35 14.95	&	-05 23 39.2	&	9.69 	&	0.9	$\pm$	0.3 	&	127 	$\pm$	10	&	194	\\
05 35 13.44	&	-05 23 40.3	&	10.09 	&	0.8	$\pm$	0.3 	&	116 	$\pm$	9	&	195	\\
05 35 11.71	&	-05 23 40.4	&	11.64 	&	0.8	$\pm$	0.4 	&	118 	$\pm$	13	&	199	\\
05 35 20.45	&	-05 23 29.8	&	9.86 	&	1.0	$\pm$	0.1 	&	116 	$\pm$	4	&	202	\\
05 35 30.16	&	-05 23 09.5	&	14.48 	&	4.8	$\pm$	0.8 	&	125 	$\pm$	5	&	216	\\
05 35 21.50	&	-05 23 16.7	&	11.20 	&	2.7	$\pm$	0.4 	&	112 	$\pm$	4	&	218	\\
05 35 09.77	&	-05 23 26.9	&	9.64 	&	1.1	$\pm$	0.1 	&	114 	$\pm$	1	&	219	\\
05 35 24.03	&	-05 23 13.9	&	12.32 	&	2.1	$\pm$	0.3 	&	118 	$\pm$	4	&	220	\\
05 35 22.80	&	-05 23 13.4	&	10.52 	&	3.8	$\pm$	0.1 	&	113 	$\pm$	1	&	221	\\
05 35 25.71	&	-05 23 09.3	&	11.04 	&	3.5	$\pm$	0.1 	&	102 	$\pm$	1	&	224	\\
05 35 08.61	&	-05 23 24.3	&	14.14 	&	4.5	$\pm$	0.9 	&	92 	$\pm$	6	&	225	\\
05 35 18.24	&	-05 23 15.6	&	10.91 	&	1.8	$\pm$	0.5 	&	127 	$\pm$	8	&	226	\\
05 35 17.82	&	-05 23 15.6	&	9.99 	&	1.2	$\pm$	0.4 	&	126 	$\pm$	10	&	227	\\
05 35 28.13	&	-05 23 06.4	&	11.03 	&	2.2	$\pm$	0.1 	&	108 	$\pm$	1	&	228	\\
05 35 21.79	&	-05 23 10.7	&	11.02 	&	2.8	$\pm$	0.3 	&	115 	$\pm$	3	&	230	\\
05 35 20.18	&	-05 23 08.6	&	11.08 	&	4.3	$\pm$	0.4 	&	110 	$\pm$	3	&	232	\\
05 35 21.84	&	-05 23 06.5	&	12.30 	&	3.5	$\pm$	1.0 	&	117 	$\pm$	8	&	233	\\
05 35 26.41	&	-05 23 02.4	&	12.44 	&	4.9	$\pm$	0.5 	&	127 	$\pm$	3	&	234	\\
05 35 10.26	&	-05 23 16.4	&	12.43 	&	7.7	$\pm$	1.6 	&	80 	$\pm$	6	&	235	\\
05 35 05.79	&	-05 23 16.0	&	12.94 	&	2.0	$\pm$	0.3 	&	109 	$\pm$	4	&	239	\\
05 35 06.43	&	-05 23 15.3	&	12.61 	&	1.8	$\pm$	0.3 	&	101 	$\pm$	5	&	240	\\
05 35 19.63	&	-05 23 03.6	&	13.67 	&	8.4	$\pm$	4.2 	&	126 	$\pm$	14	&	241	\\
05 35 14.31	&	-05 23 08.3	&	11.16 	&	5.2	$\pm$	0.5 	&	131 	$\pm$	3	&	242	\\
05 35 26.16	&	-05 22 57.0	&	11.81 	&	2.2	$\pm$	0.4 	&	147 	$\pm$	5	&	243	\\
05 35 21.24	&	-05 22 59.5	&	11.54 	&	2.3	$\pm$	0.3 	&	115 	$\pm$	4	&	245	\\
05 35 14.87	&	-05 23 05.1	&	11.08 	&	8.9	$\pm$	1.2 	&	132 	$\pm$	4	&	246	\\
05 35 27.78	&	-05 22 52.3	&	14.64 	&	3.5	$\pm$	1.6 	&	6 	$\pm$	13	&	247	\\
05 35 14.66	&	-05 23 01.9	&	11.41 	&	4.6	$\pm$	1.4 	&	139 	$\pm$	9	&	252	\\
05 35 20.63	&	-05 22 55.7	&	12.73 	&	2.8	$\pm$	1.1 	&	125 	$\pm$	11	&	253	\\
05 35 12.57	&	-05 23 02.0	&	11.81 	&	14.7	$\pm$	1.2 	&	130 	$\pm$	2	&	255	\\
05 35 08.43	&	-05 23 05.0	&	12.99 	&	5.7	$\pm$	0.8 	&	120 	$\pm$	4	&	256	\\
05 35 27.96	&	-05 22 47.0	&	13.75 	&	6.7	$\pm$	0.6 	&	111 	$\pm$	2	&	257	\\
05 35 14.36	&	-05 22 54.1	&	11.41 	&	6.6	$\pm$	0.9 	&	126 	$\pm$	4	&	260	\\
05 35 10.61	&	-05 22 56.1	&	11.52 	&	21.0	$\pm$	0.6 	&	119 	$\pm$	1	&	261	\\
05 35 11.97	&	-05 22 54.2	&	10.73 	&	18.0	$\pm$	0.5 	&	128 	$\pm$	1	&	263	\\
05 35 08.73	&	-05 22 56.7	&	11.11 	&	8.5	$\pm$	0.2 	&	118 	$\pm$	1	&	265	\\
05 35 24.66	&	-05 22 42.6	&	11.54 	&	1.9	$\pm$	0.2 	&	108 	$\pm$	3	&	266	\\
05 35 14.70	&	-05 22 49.4	&	12.12 	&	5.3	$\pm$	1.7 	&	12 	$\pm$	9	&	269	\\
05 35 15.48	&	-05 22 48.6	&	10.19 	&	1.1	$\pm$	0.4 	&	157 	$\pm$	11	&	270	\\
05 35 22.97	&	-05 22 41.6	&	10.99 	&	1.6	$\pm$	0.1 	&	140 	$\pm$	2	&	271	\\
05 35 10.50	&	-05 22 45.7	&	9.87 	&	1.0	$\pm$	0.1 	&	120 	$\pm$	4	&	274	\\
05 35 22.12	&	-05 22 34.1	&	11.46 	&	3.1	$\pm$	0.8 	&	145 	$\pm$	7	&	276	\\
05 35 14.09	&	-05 22 36.6	&	10.11 	&	1.3	$\pm$	0.3 	&	133 	$\pm$	6	&	284	\\
05 35 23.18	&	-05 22 28.4	&	11.99 	&	1.9	$\pm$	0.4 	&	137 	$\pm$	6	&	285	\\
05 35 17.01	&	-05 22 33.1	&	9.88 	&	0.8	$\pm$	0.2 	&	123 	$\pm$	8	&	287	\\
05 35 11.20	&	-05 22 37.8	&	11.54 	&	25.2	$\pm$	0.5 	&	128 	$\pm$	1	&	288	\\
05 35 25.13	&	-05 22 25.2	&	12.04 	&	1.2	$\pm$	0.4 	&	126 	$\pm$	9	&	289	\\
05 35 15.88	&	-05 22 33.2	&	12.30 	&	14.9	$\pm$	2.4 	&	5 	$\pm$	5	&	290	\\
05 35 22.84	&	-05 22 27.0	&	12.60 	&	1.8	$\pm$	0.7 	&	131 	$\pm$	12	&	291	\\
05 35 17.75	&	-05 22 31.0	&	12.03 	&	10.7	$\pm$	0.9 	&	113 	$\pm$	3	&	292	\\
05 35 21.03	&	-05 22 25.2	&	12.03 	&	2.8	$\pm$	0.5 	&	172 	$\pm$	5	&	298	\\
05 35 04.82	&	-05 22 38.8	&	12.01 	&	0.4	$\pm$	0.2 	&	70 	$\pm$	12	&	299	\\
05 35 10.14	&	-05 22 32.7	&	11.53 	&	2.4	$\pm$	0.5 	&	83 	$\pm$	6	&	300	\\
05 35 19.82	&	-05 22 21.6	&	10.68 	&	3.1	$\pm$	0.5 	&	117 	$\pm$	4	&	302	\\
05 35 08.41	&	-05 22 30.2	&	12.77 	&	12.9	$\pm$	0.7 	&	128 	$\pm$	2	&	304	\\
05 35 16.89	&	-05 22 22.5	&	9.78 	&	1.1	$\pm$	0.2 	&	123 	$\pm$	5	&	305	\\
05 35 18.95	&	-05 22 18.8	&	10.52 	&	4.1	$\pm$	0.3 	&	110 	$\pm$	2	&	308	\\
05 35 05.45	&	-05 22 30.5	&	13.85 	&	4.1	$\pm$	0.5 	&	123 	$\pm$	3	&	309	\\
05 35 29.77	&	-05 22 08.5	&	14.36 	&	7.6	$\pm$	1.2 	&	169 	$\pm$	4	&	311	\\
05 35 10.99	&	-05 22 24.8	&	11.88 	&	3.2	$\pm$	0.5 	&	119 	$\pm$	4	&	312	\\
05 35 13.74	&	-05 22 22.0	&	10.16 	&	1.7	$\pm$	0.6 	&	118 	$\pm$	10	&	314	\\
05 35 07.27	&	-05 22 26.6	&	11.42 	&	1.2	$\pm$	0.2 	&	112 	$\pm$	4	&	316	\\
05 35 13.17	&	-05 22 21.1	&	11.17 	&	1.7	$\pm$	0.6 	&	129 	$\pm$	10	&	318	\\
05 35 20.40	&	-05 22 13.7	&	10.04 	&	0.4	$\pm$	0.1 	&	138 	$\pm$	10	&	319	\\
05 35 10.71	&	-05 22 20.4	&	14.26 	&	16.2	$\pm$	5.5 	&	158 	$\pm$	10	&	321	\\
05 35 30.99	&	-05 22 01.3	&	11.40 	&	7.1	$\pm$	0.1 	&	127 	$\pm$	1	&	322	\\
05 35 04.17	&	-05 22 24.4	&	12.89 	&	1.0	$\pm$	0.2 	&	106 	$\pm$	5	&	323	\\
05 35 10.53	&	-05 22 16.6	&	11.54 	&	5.9	$\pm$	0.4 	&	135 	$\pm$	2	&	327	\\
05 35 18.03	&	-05 22 05.4	&	10.35 	&	4.9	$\pm$	0.3 	&	104 	$\pm$	2	&	331	\\
05 35 22.45	&	-05 22 01.1	&	10.53 	&	1.0	$\pm$	0.2 	&	111 	$\pm$	5	&	332	\\
05 35 28.63	&	-05 21 54.0	&	13.84 	&	1.8	$\pm$	0.6 	&	140 	$\pm$	10	&	335	\\
05 35 21.22	&	-05 22 00.2	&	11.33 	&	3.1	$\pm$	0.8 	&	96 	$\pm$	8	&	336	\\
05 35 18.76	&	-05 22 02.2	&	10.67 	&	1.2	$\pm$	0.2 	&	159 	$\pm$	6	&	337	\\
05 35 17.88	&	-05 22 02.9	&	12.68 	&	19.1	$\pm$	2.4 	&	94 	$\pm$	4	&	338	\\
05 35 06.18	&	-05 22 12.5	&	11.79 	&	7.4	$\pm$	0.2 	&	126 	$\pm$	1	&	339	\\
05 35 24.12	&	-05 21 55.6	&	12.49 	&	1.8	$\pm$	0.4 	&	112 	$\pm$	6	&	342	\\
05 35 25.43	&	-05 21 51.6	&	10.69 	&	2.2	$\pm$	0.1 	&	128 	$\pm$	2	&	344	\\
05 35 30.74	&	-05 21 46.7	&	12.40 	&	7.3	$\pm$	0.3 	&	131 	$\pm$	1	&	345	\\
05 35 20.77	&	-05 21 55.1	&	10.50 	&	1.0	$\pm$	0.3 	&	153 	$\pm$	10	&	347	\\
05 35 06.45	&	-05 22 07.6	&	12.75 	&	1.1	$\pm$	0.5 	&	120 	$\pm$	13	&	348	\\
05 35 14.99	&	-05 21 59.9	&	10.03 	&	0.5	$\pm$	0.2 	&	112 	$\pm$	12	&	349	\\
05 35 19.96	&	-05 21 53.9	&	13.53 	&	21.9	$\pm$	2.6 	&	96 	$\pm$	3	&	351	\\
05 35 20.93	&	-05 21 50.9	&	10.39 	&	0.9	$\pm$	0.4 	&	128 	$\pm$	12	&	352	\\
05 35 17.61	&	-05 21 53.9	&	11.84 	&	6.2	$\pm$	0.9 	&	130 	$\pm$	4	&	353	\\
05 35 15.26	&	-05 21 55.7	&	12.50 	&	11.0	$\pm$	2.1 	&	111 	$\pm$	5	&	354	\\
05 35 06.28	&	-05 22 02.7	&	9.88 	&	14.2	$\pm$	0.1 	&	129 	$\pm$	1	&	355	\\
05 35 19.54	&	-05 21 49.1	&	13.92 	&	8.9	$\pm$	2.7 	&	110 	$\pm$	9	&	356	\\
05 35 21.67	&	-05 21 47.1	&	10.77 	&	1.1	$\pm$	0.2 	&	119 	$\pm$	4	&	357	\\
05 35 30.60	&	-05 21 39.1	&	13.64 	&	14.5	$\pm$	0.5 	&	118 	$\pm$	1	&	359	\\
05 35 10.58	&	-05 21 56.3	&	10.19 	&	1.2	$\pm$	0.1 	&	132 	$\pm$	3	&	361	\\
05 35 13.06	&	-05 21 53.3	&	12.05 	&	7.5	$\pm$	2.0 	&	134 	$\pm$	8	&	363	\\
05 35 31.49	&	-05 21 36.5	&	13.09 	&	7.0	$\pm$	0.3 	&	125 	$\pm$	1	&	364	\\
05 35 22.28	&	-05 21 42.6	&	13.67 	&	4.7	$\pm$	2.1 	&	128 	$\pm$	13	&	365	\\
05 35 16.00	&	-05 21 47.1	&	11.76 	&	1.1	$\pm$	0.5 	&	134 	$\pm$	12	&	367	\\
05 35 17.55	&	-05 21 45.6	&	10.52 	&	1.4	$\pm$	0.3 	&	143 	$\pm$	6	&	368	\\
05 35 19.16	&	-05 21 43.6	&	13.16 	&	6.1	$\pm$	2.0 	&	107 	$\pm$	9	&	370	\\
05 35 27.33	&	-05 21 35.8	&	13.72 	&	7.5	$\pm$	0.8 	&	107 	$\pm$	3	&	371	\\
05 35 11.80	&	-05 21 49.3	&	11.91 	&	27.8	$\pm$	0.9 	&	135 	$\pm$	1	&	372	\\
05 35 28.16	&	-05 21 34.8	&	12.87 	&	8.2	$\pm$	0.2 	&	120 	$\pm$	1	&	373	\\
05 35 18.85	&	-05 21 41.2	&	10.57 	&	0.9	$\pm$	0.3 	&	127 	$\pm$	8	&	374	\\
05 35 29.04	&	-05 21 31.9	&	14.54 	&	5.6	$\pm$	0.9 	&	128 	$\pm$	4	&	375	\\
05 35 09.52	&	-05 21 48.1	&	14.31 	&	13.3	$\pm$	4.3 	&	141 	$\pm$	9	&	377	\\
05 35 23.21	&	-05 21 35.9	&	12.12 	&	1.6	$\pm$	0.6 	&	138 	$\pm$	11	&	378	\\
05 35 24.09	&	-05 21 32.7	&	11.45 	&	1.6	$\pm$	0.2 	&	128 	$\pm$	3	&	380	\\
05 35 15.76	&	-05 21 39.9	&	11.84 	&	1.0	$\pm$	0.4 	&	161 	$\pm$	12	&	381	\\
05 35 15.40	&	-05 21 39.5	&	13.42 	&	11.1	$\pm$	2.0 	&	137 	$\pm$	5	&	382	\\
05 35 09.92	&	-05 21 43.4	&	12.21 	&	11.3	$\pm$	0.4 	&	132 	$\pm$	1	&	383	\\
05 35 20.13	&	-05 21 33.7	&	10.01 	&	0.8	$\pm$	0.2 	&	129 	$\pm$	6	&	384	\\
05 35 09.34	&	-05 21 41.6	&	11.53 	&	7.5	$\pm$	0.2 	&	137 	$\pm$	1	&	385	\\
05 35 20.84	&	-05 21 30.1	&	11.26 	&	1.3	$\pm$	0.3 	&	108 	$\pm$	6	&	386	\\
05 35 13.72	&	-05 21 35.9	&	13.04 	&	22.2	$\pm$	1.7 	&	133 	$\pm$	2	&	387	\\
05 35 23.33	&	-05 21 25.4	&	11.26 	&	2.6	$\pm$	0.3 	&	101 	$\pm$	3	&	389	\\
05 35 12.86	&	-05 21 33.9	&	12.18 	&	26.8	$\pm$	1.4 	&	139 	$\pm$	2	&	391	\\
05 35 27.76	&	-05 21 18.9	&	13.35 	&	12.7	$\pm$	0.7 	&	112 	$\pm$	2	&	393	\\
05 35 11.76	&	-05 21 32.8	&	13.18 	&	8.0	$\pm$	2.1 	&	122 	$\pm$	7	&	394	\\
05 35 15.30	&	-05 21 28.9	&	14.50 	&	14.3	$\pm$	3.4 	&	125 	$\pm$	7	&	395	\\
05 35 20.84	&	-05 21 21.6	&	11.55 	&	1.7	$\pm$	0.3 	&	46 	$\pm$	6	&	396	\\
05 35 23.88	&	-05 21 18.6	&	12.33 	&	3.8	$\pm$	0.6 	&	82 	$\pm$	4	&	397	\\
05 35 23.66	&	-05 21 16.4	&	13.73 	&	7.5	$\pm$	2.3 	&	83 	$\pm$	9	&	399	\\
05 35 09.77	&	-05 21 28.4	&	10.06 	&	8.7	$\pm$	0.1 	&	141 	$\pm$	1	&	400	\\
05 35 13.97	&	-05 21 23.3	&	10.94 	&	11.2	$\pm$	0.2 	&	143 	$\pm$	1	&	401	\\
05 35 15.40	&	-05 21 14.0	&	10.95 	&	1.7	$\pm$	0.2 	&	130 	$\pm$	4	&	407	\\
05 35 08.42	&	-05 21 19.8	&	13.55 	&	6.8	$\pm$	0.8 	&	143 	$\pm$	3	&	408	\\
05 35 24.62	&	-05 21 04.3	&	12.95 	&	7.9	$\pm$	0.4 	&	95 	$\pm$	1	&	409	\\
05 35 11.31	&	-05 21 15.5	&	13.96 	&	19.6	$\pm$	2.0 	&	145 	$\pm$	3	&	410	\\
05 35 16.19	&	-05 21 10.9	&	11.70 	&	4.1	$\pm$	0.3 	&	26 	$\pm$	2	&	412	\\
05 35 21.55	&	-05 21 05.7	&	11.13 	&	1.9	$\pm$	0.3 	&	115 	$\pm$	4	&	414	\\
05 35 18.96	&	-05 21 07.8	&	10.48 	&	0.8	$\pm$	0.1 	&	114 	$\pm$	2	&	415	\\
05 35 06.46	&	-05 21 18.8	&	12.78 	&	1.1	$\pm$	0.3 	&	131 	$\pm$	9	&	416	\\
05 35 07.95	&	-05 21 17.2	&	12.60 	&	1.1	$\pm$	0.3 	&	120 	$\pm$	7	&	417	\\
05 35 16.28	&	-05 21 09.2	&	10.83 	&	2.1	$\pm$	0.1 	&	153 	$\pm$	2	&	419	\\
05 35 10.32	&	-05 21 13.1	&	10.53 	&	4.3	$\pm$	0.1 	&	106 	$\pm$	1	&	420	\\
05 35 19.50	&	-05 21 04.6	&	12.29 	&	1.5	$\pm$	0.5 	&	115 	$\pm$	10	&	421	\\
05 35 23.43	&	-05 21 00.0	&	13.90 	&	2.7	$\pm$	0.8 	&	109 	$\pm$	8	&	422	\\
05 35 14.71	&	-05 21 06.3	&	11.48 	&	12.0	$\pm$	0.2 	&	138 	$\pm$	1	&	423	\\
05 35 13.43	&	-05 21 07.3	&	11.98 	&	8.7	$\pm$	0.4 	&	129 	$\pm$	1	&	424	\\
05 35 26.76	&	-05 20 53.2	&	13.62 	&	3.6	$\pm$	0.7 	&	134 	$\pm$	5	&	425	\\
05 35 12.85	&	-05 21 05.0	&	13.78 	&	7.4	$\pm$	1.7 	&	121 	$\pm$	7	&	426	\\
05 35 16.94	&	-05 20 59.9	&	12.00 	&	2.2	$\pm$	0.7 	&	107 	$\pm$	9	&	427	\\
05 35 14.95	&	-05 21 00.8	&	13.59 	&	6.1	$\pm$	2.4 	&	114 	$\pm$	11	&	428	\\
05 35 11.89	&	-05 21 03.3	&	10.38 	&	10.2	$\pm$	0.1 	&	128 	$\pm$	1	&	429	\\
05 35 04.95	&	-05 21 09.3	&	12.16 	&	8.2	$\pm$	0.1 	&	154 	$\pm$	1	&	430	\\
05 35 11.28	&	-05 21 03.0	&	13.30 	&	11.6	$\pm$	1.4 	&	135 	$\pm$	4	&	431	\\
05 35 20.52	&	-05 20 52.2	&	11.14 	&	1.6	$\pm$	0.3 	&	147 	$\pm$	6	&	432	\\
05 35 17.85	&	-05 20 54.1	&	10.02 	&	2.2	$\pm$	0.1 	&	118 	$\pm$	1	&	433	\\
05 35 18.95	&	-05 20 52.3	&	11.38 	&	2.5	$\pm$	0.3 	&	116 	$\pm$	3	&	435	\\
05 35 07.74	&	-05 21 01.5	&	11.95 	&	7.1	$\pm$	0.2 	&	123 	$\pm$	1	&	436	\\
05 35 13.21	&	-05 20 52.8	&	11.25 	&	2.2	$\pm$	0.2 	&	118 	$\pm$	3	&	440	\\
05 35 04.87	&	-05 20 57.6	&	11.92 	&	6.9	$\pm$	0.1 	&	140 	$\pm$	1	&	441	\\
05 35 21.00	&	-05 20 43.2	&	13.65 	&	6.3	$\pm$	0.7 	&	115 	$\pm$	3	&	442	\\
05 35 20.08	&	-05 20 44.0	&	11.14 	&	4.8	$\pm$	0.2 	&	148 	$\pm$	1	&	443	\\
05 35 20.57	&	-05 20 43.3	&	11.31 	&	1.7	$\pm$	0.1 	&	133 	$\pm$	2	&	444	\\
05 35 18.52	&	-05 20 42.8	&	11.50 	&	1.9	$\pm$	0.2 	&	130 	$\pm$	3	&	445	\\
05 35 12.39	&	-05 20 47.9	&	12.41 	&	4.7	$\pm$	0.7 	&	121 	$\pm$	4	&	446	\\
05 35 25.00	&	-05 20 34.3	&	14.52 	&	21.5	$\pm$	1.6 	&	135 	$\pm$	2	&	447	\\
05 35 12.27	&	-05 20 45.2	&	10.93 	&	2.8	$\pm$	0.1 	&	113 	$\pm$	1	&	448	\\
05 35 12.82	&	-05 20 43.7	&	10.47 	&	1.2	$\pm$	0.1 	&	8 	$\pm$	3	&	452	\\
05 35 15.87	&	-05 20 40.5	&	11.81 	&	1.8	$\pm$	0.2 	&	120 	$\pm$	4	&	453	\\
05 35 08.23	&	-05 20 47.0	&	12.84 	&	8.5	$\pm$	0.5 	&	122 	$\pm$	2	&	454	\\
05 35 13.58	&	-05 20 39.2	&	13.35 	&	9.6	$\pm$	1.0 	&	153 	$\pm$	3	&	456	\\
05 35 16.07	&	-05 20 36.3	&	10.00 	&	1.8	$\pm$	0.1 	&	115 	$\pm$	1	&	457	\\
05 35 12.10	&	-05 20 39.8	&	14.25 	&	7.4	$\pm$	1.8 	&	126 	$\pm$	7	&	458	\\
05 35 12.77	&	-05 20 34.9	&	12.22 	&	2.0	$\pm$	0.5 	&	136 	$\pm$	7	&	461	\\
05 35 13.59	&	-05 20 31.4	&	12.76 	&	3.9	$\pm$	0.5 	&	135 	$\pm$	4	&	463	\\
05 35 17.23	&	-05 20 27.8	&	10.82 	&	1.2	$\pm$	0.3 	&	104 	$\pm$	6	&	464	\\
05 35 15.27	&	-05 20 29.0	&	13.03 	&	1.2	$\pm$	0.6 	&	105 	$\pm$	14	&	465	\\
05 35 20.18	&	-05 20 24.7	&	12.89 	&	5.3	$\pm$	0.6 	&	128 	$\pm$	3	&	466	\\
05 35 14.76	&	-05 20 29.0	&	10.84 	&	7.9	$\pm$	0.1 	&	141 	$\pm$	1	&	467	\\
05 35 13.04	&	-05 20 30.3	&	10.89 	&	4.5	$\pm$	0.1 	&	139 	$\pm$	1	&	468	\\
05 35 16.31	&	-05 20 25.3	&	12.11 	&	0.9	$\pm$	0.4 	&	141 	$\pm$	14	&	469	\\
05 35 18.39	&	-05 20 20.4	&	11.17 	&	4.8	$\pm$	0.1 	&	140 	$\pm$	1	&	470	\\
05 35 14.16	&	-05 20 23.5	&	13.66 	&	6.5	$\pm$	1.1 	&	137 	$\pm$	5	&	471	\\
05 35 16.73	&	-05 20 20.0	&	12.18 	&	2.5	$\pm$	0.8 	&	151 	$\pm$	10	&	473	\\
05 35 18.36	&	-05 20 16.5	&	11.46 	&	1.7	$\pm$	0.1 	&	142 	$\pm$	2	&	474	\\
05 35 13.31	&	-05 20 18.9	&	10.12 	&	6.0	$\pm$	0.1 	&	142 	$\pm$	1	&	475	\\
05 35 17.37	&	-05 20 14.9	&	11.38 	&	1.6	$\pm$	0.3 	&	141 	$\pm$	5	&	476	\\
05 35 10.19	&	-05 20 21.0	&	12.46 	&	2.0	$\pm$	0.3 	&	120 	$\pm$	4	&	477	\\
05 35 26.17	&	-05 20 06.1	&	10.54 	&	7.4	$\pm$	0.1 	&	132 	$\pm$	1	&	478	\\
05 35 05.13	&	-05 20 24.5	&	11.29 	&	1.8	$\pm$	0.1 	&	120 	$\pm$	1	&	479	\\
05 35 15.21	&	-05 20 15.0	&	13.27 	&	1.9	$\pm$	0.8 	&	136 	$\pm$	13	&	480	\\
05 35 29.64	&	-05 20 02.2	&	12.62 	&	0.3	$\pm$	0.1 	&	147 	$\pm$	13	&	481	\\
05 35 17.02	&	-05 20 12.1	&	12.67 	&	2.9	$\pm$	1.2 	&	130 	$\pm$	12	&	482	\\
05 35 21.81	&	-05 20 07.8	&	14.80 	&	13.1	$\pm$	2.0 	&	129 	$\pm$	4	&	483	\\
05 35 06.78	&	-05 20 19.1	&	13.66 	&	2.9	$\pm$	0.4 	&	112 	$\pm$	4	&	485	\\
05 35 06.19	&	-05 20 17.4	&	13.20 	&	7.8	$\pm$	0.3 	&	145 	$\pm$	1	&	487	\\
05 35 23.49	&	-05 20 01.8	&	12.05 	&	12.5	$\pm$	0.4 	&	127 	$\pm$	1	&	488	\\
05 35 09.02	&	-05 20 11.8	&	13.19 	&	6.6	$\pm$	0.5 	&	169 	$\pm$	2	&	489	\\
05 35 04.16	&	-05 20 15.8	&	12.23 	&	5.8	$\pm$	0.2 	&	136 	$\pm$	1	&	490	\\
05 35 19.61	&	-05 20 01.9	&	12.60 	&	9.3	$\pm$	0.6 	&	131 	$\pm$	2	&	491	\\
05 35 04.55	&	-05 20 13.9	&	13.59 	&	8.8	$\pm$	0.3 	&	145 	$\pm$	1	&	492	\\
05 35 14.22	&	-05 20 04.3	&	11.59 	&	5.8	$\pm$	0.1 	&	135 	$\pm$	1	&	493	\\
05 35 20.09	&	-05 19 59.1	&	12.27 	&	6.0	$\pm$	0.4 	&	147 	$\pm$	2	&	494	\\
05 35 04.19	&	-05 20 12.4	&	13.80 	&	1.9	$\pm$	0.8 	&	114 	$\pm$	13	&	495	\\
05 35 20.31	&	-05 19 58.0	&	14.07 	&	13.1	$\pm$	2.1 	&	129 	$\pm$	5	&	496	\\
05 35 11.89	&	-05 20 02.0	&	13.21 	&	6.6	$\pm$	0.7 	&	143 	$\pm$	3	&	497	\\
05 35 07.97	&	-05 20 03.4	&	14.51 	&	14.1	$\pm$	1.1 	&	143 	$\pm$	2	&	498	\\
\enddata
\\
Note (1): All H band magnitude errors are less than 0.1 mag.\\
Note (2): This list picked up the detected polarized sources 
(polarization S/N  2) in H band from all point sources which 
were detected in the field. The all sources will be published 
in future paper.
%
\end{deluxetable}
\end{center}

\begin{center}
\begin{table}[t]
\caption{ Summary of polarization statistics.}
\begin{tabular}{lccc} \hline
 band & $< \theta >$ & $s$ & number\\ \hline
 $J$ & 120$\pm$2 & 0.35$\pm$0.03 & 200 \\
 $H$ & 119$\pm$1 & 0.25$\pm$0.02 & 313\\
 $Ks$ & 124$\pm$1 & 0.25$\pm$0.01 & 278 \\
 350 $\mu$m & 120$\pm$2 & 0.43$\pm$0.03 & 470 \\ \hline
 \label{PA}
\end{tabular}
\end{table}
\end{center}


\begin{center}
\begin{table}
\caption{ $H$ band polarimetry of SHC sources (Slesnick et al. 2004).}
\begin{tabular}{cccccccc} \hline
name	&	RA (J2000)	&	Dec (J2000)	&	$H$	&	$A(V)$	&	Mass		&	$P(H)$	&	$P(H)/A(H)$ \\
	&	(h:m:s)	&	(d:m:s)	&	(mag)	&	(mag)	&	($M_\odot $)	&	(\%)	&	(\%/mag)	\\ \hline
[HC2000]	&		&	&	&	&		&		&		\\
20	&	05 35 21.35	&	-05 25 35.0	&	14.87	&	3.77	&	0.04	&	0	$\pm$	4	&	0	\\
22	&	05 35 20.90	&	-05 25 34.5	&	13.74	&	1.78	&	0.05	&	1.5	$\pm$	1.7	&	0	\\
25	&	05 35 20.16	&	-05 25 33.9	&	14.47	&	6	&	0.42	&	1.7	$\pm$	2.7	&	1.6	\\
29	&	05 35 25.35	&	-05 25 29.5	&	12.37	&	4.77	&	0.11	&	0.6	$\pm$	0.2	&	0.7	\\
35	&	05 35 17.95	&	-05 25 21.3	&	11.64	&	1.94	&	0.1	&	1.7	$\pm$	1.3	&	5	\\
51	&	05 35 24.89	&	-05 25 10.2	&	12.10	&	2.26	&	0.11	&	1.1	$\pm$	0.3	&	2.7	\\
64	&	05 35 07.06	&	-05 25 00.9	&	15.25	&	4.46	&	0.03	&	5.8	$\pm$	4.5	&	4.7	\\
90	&	05 35 10.38	&	-05 24 51.6	&	13.87	&	6.77	&	0.07	&	2.9	$\pm$	1.3	&	2.4	\\
93	&	05 35 10.95	&	-05 24 48.8	&	10.27	&	8.24	&	0.18	&	1	$\pm$	0.1	&	0.7	\\
114	&	05 35 25.03	&	-05 24 38.4	&	14.44	&	1.44	&	0.05	&	0	$\pm$	2.2	&	0	\\
264	&	05 35 21.18	&	-05 23 33.1	&	12.87	&	1.67	&	0.1	&	0	$\pm$	1.2	&	0	\\
288	&	05 35 19.12	&	-05 23 27.1	&	11.06	&	4.08	&	0.23	&	0.6	$\pm$	0.4	&	0.9	\\
290	&	05 35 08.62	&	-05 23 24.4	&	14.24	&	6.64	&	0.11	&	4.5	$\pm$	0.9	&	3.8	\\
346	&	05 35 20.18	&	-05 23 08.5	&	11.06	&	5.81	&	0.18	&	4.3	$\pm$	0.4	&	4.2	\\
355	&	05 35 21.89	&	-05 23 07.2	&	12.88	&	3.32	&	0.12	&	3.5	$\pm$	1	&	5.9	\\
366	&	05 35 19.63	&	-05 23 03.6	&	13.77	&	10.25	&	0.08	&	8.4	$\pm$	4.2	&	4.6	\\
368	&	05 35 12.58	&	-05 23 02.0	&	11.95	&	8.77	&	0.1	&	14.7	$\pm$	1.2	&	9.4	\\
381	&	05 35 21.25	&	-05 22 59.5	&	11.54	&	3.93	&	0.17	&	2.3	$\pm$	0.3	&	3.3	\\
395	&	05 35 11.98	&	-05 22 54.2	&	10.84	&	12.41	&	0.2	&	18	$\pm$	0.5	&	8.1	\\
396	&	05 35 20.63	&	-05 22 55.7	&	13.17	&	13.53	&	0.1	&	2.8	$\pm$	1.1	&	1.2	\\
434	&	05 35 11.20	&	-05 22 37.8	&	12.64	&	21.41	&	0.2	&	25.2	$\pm$	0.5	&	6.6	\\
454	&	05 35 08.42	&	-05 22 30.3	&	12.78	&	7.85	&	0.23	&	12.9	$\pm$	0.7	&	9.2	\\
459	&	05 35 07.39	&	-05 22 29.0	&	12.80	&	1.59	&	0.1	&	0.1	$\pm$	0.5	&	0.2	\\
467	&	05 35 07.28	&	-05 22 26.6	&	11.69	&	2.55	&	0.1	&	1.2	$\pm$	0.2	&	2.6	\\
543	&	05 35 18.76	&	-05 22 02.2	&	10.71	&	4	&	0.21	&	1.2	$\pm$	0.2	&	1.6	\\
553	&	05 35 10.27	&	-05 21 57.2	&	13.00	&	5.77	&	0.11	&	0	$\pm$	1	&	0	\\
555	&	05 35 11.79	&	-05 21 55.6	&	12.04	&	1.56	&	0.1	&	0.6	$\pm$	0.7	&	2.1	\\
600	&	05 35 15.41	&	-05 21 39.5	&	13.50	&	14.26	&	0.11	&	11.1	$\pm$	2	&	4.4	\\

\hline
 \label{AvMP_S}
\end{tabular}
Note.-- The IDs from Hillenbrand \& Carpenter (2000).

\end{table}
\end{center}


\begin{center}
\begin{table}
\caption{ $H$ band polarimetry of Riddick sources (Riddick al. 2007).}
\begin{tabular}{cccccccc} \hline
[OW94]	&	RA (J2000)	&	Dec (J2000)	&	$H$	&	$A_V$	&	Mass &	$P(H)$  &	P(H)/A(H)\\
	&	(h:m:s)	&	(d:m:s)	&	(mag)	&	(mag)	&	($M_\odot $) 	&	(\%)	&	(\%/mag)	\\
112-532	&	5	35	11.15	&	-5	25	32	&	12.98 	&	3.6	&	0.17	&	0.0 	$\pm$	0.5 	&	0.0 	\\
082-403	&	5	35	8.23	&	-5	24	3.2	&	12.19 	&	4.1	&	0.17	&	2.7 	$\pm$	0.3 	&	3.8 	\\
130-458	&	5	35	12.96	&	-5	24	58	&	12.95 	&	4.0	&	0.14	&	0.0 	$\pm$	2.6 	&	0.0 	\\
121-434	&	5	35	12.12	&	-5	24	34	&	13.28 	&	4.4	&	0.12	&	4.1 	$\pm$	2.2 	&	5.4 	\\
068-019	&	5	35	6.76	&	-5	20	19	&	13.66 	&	0.6	&	0.11	&	2.9 	$\pm$	0.4 	&	27.3 	\\
103-157	&	5	35	10.28	&	-5	21	57	&	12.46 	&	1.0	&	0.11	&	0.0 	$\pm$	1.0 	&	0.0 	\\
077-453	&	5	35	7.7	&	-5	24	52.5	&	13.08 	&	1.9	&	0.076	&	1.3 	$\pm$	0.6 	&	4.0 	\\
072-638	&	5	35	7.21	&	-5	26	38.2	&	14.65 	&	1.1	&	0.076	&	1.9 	$\pm$	1.0 	&	10.0 	\\
053-503	&	5	35	5.28	&	-5	25	2.8	&	13.35 	&	2.1	&	0.076	&	2.8 	$\pm$	0.6 	&	7.6 	\\
154-600	&	5	35	15.42	&	-5	25	59.6	&	12.85 	&	0.9	&	0.076	&	0.6 	$\pm$	1.3 	&	3.8 	\\
055-230	&	5	35	5.47	&	-5	22	30.3	&	13.85 	&	0.9	&	0.066	&	4.1 	$\pm$	0.5 	&	26.0 	\\
084-305	&	5	35	8.44	&	-5	23	4.9	&	12.99 	&	4.0	&	0.058	&	5.7 	$\pm$	0.8 	&	8.1 	\\
042-012	&	5	35	4.18	&	-5	20	12.3	&	13.80 	&	0.9	&	0.036	&	1.9 	$\pm$	0.8 	&	12.3 	\\
186-631	&	5	35	18.61	&	-5	26	31.4	&	15.07 	&	1.2	&	0.031	&	0.0 	$\pm$	4.1 	&	0.0 	\\
077-127	&	5	35	7.74	&	-5	21	26.6	&	14.01 	&	0.9	&	0.021	&	1.0 	$\pm$	1.0 	&	6.6 	\\

\hline
 \label{AvMP_R}
\end{tabular}
Note.-- The names from O'Dell \& Wen (1994).
\end{table}
\end{center}

\clearpage

\clearpage
\begin{figure}[htb]
\includegraphics[width=9cm]{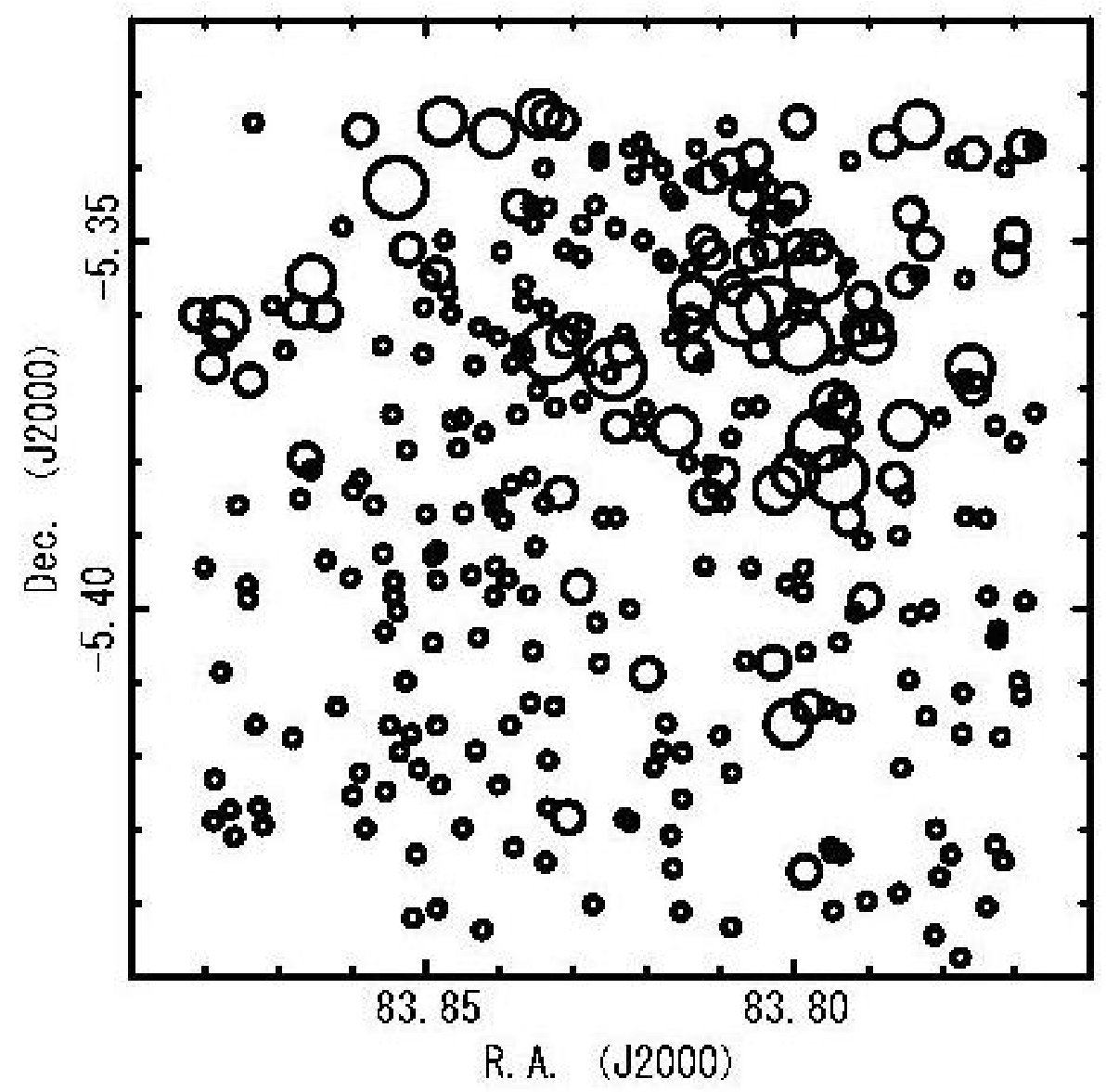}
\caption{Spatial distribution of $H$ band polarization degrees.
Each circle represents $P$ $<$ 6\%, 6\% $<$ $P$ $<$ 12\%, 12\% $<$ $P$ $<$ 18\%, or 18\% $<$ $P$ .}
\label{hpsmap}
\end{figure}

\clearpage
\begin{figure}[htb]
\includegraphics[width=15cm]{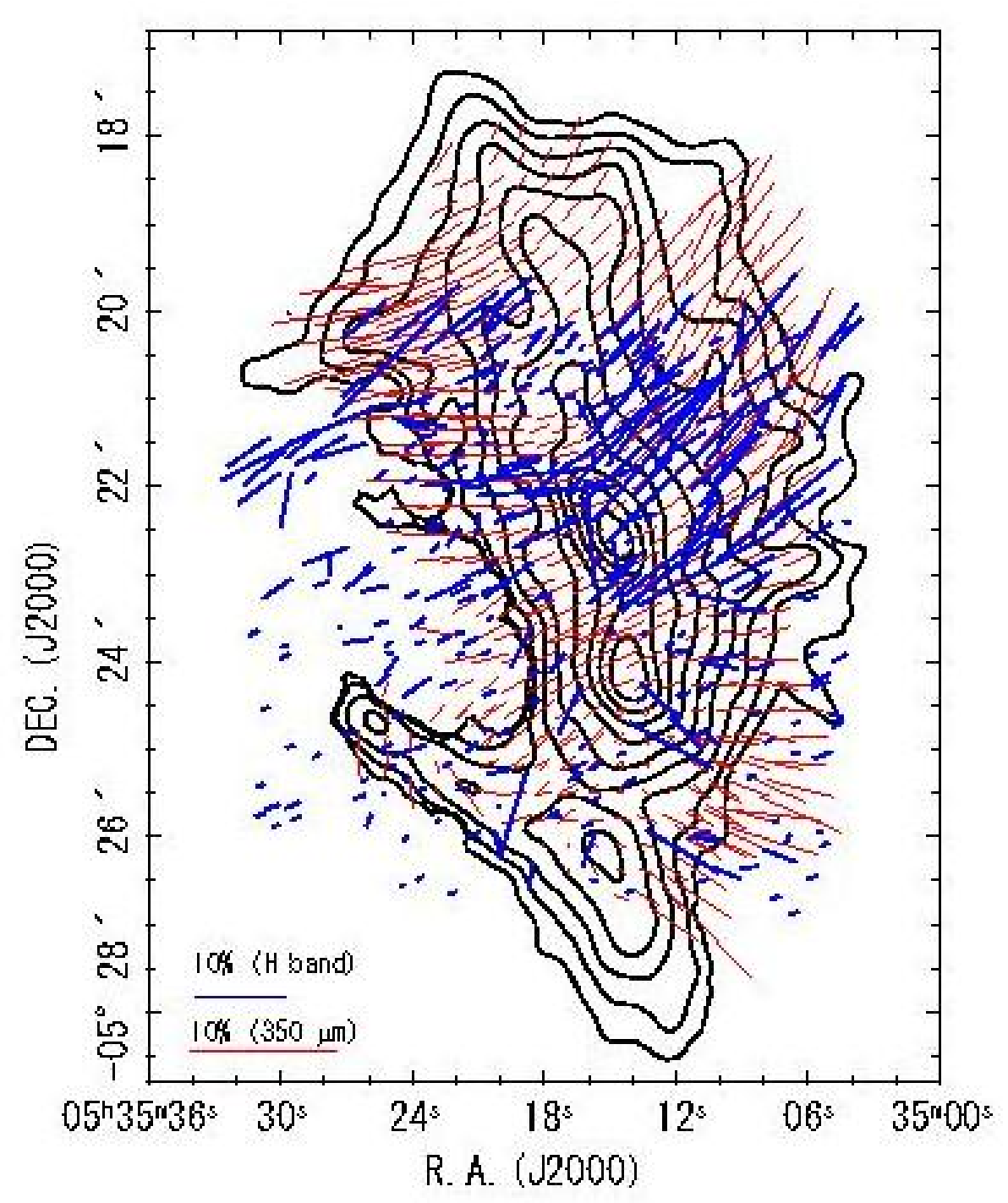}
\caption{$H$ band (shown in blue) polarization vectors and 350 $\mu$m 
(shown in red, rotated by 90$^\circ$) 
polarization vectors overlaid on the 350 $\mu$m contours.
10\% polarization vectors for both wavelengths are shown at the bottom-left.
}
\label{nir_sm}
\end{figure}

\clearpage
\begin{figure}[htb]
\includegraphics[width=16cm]{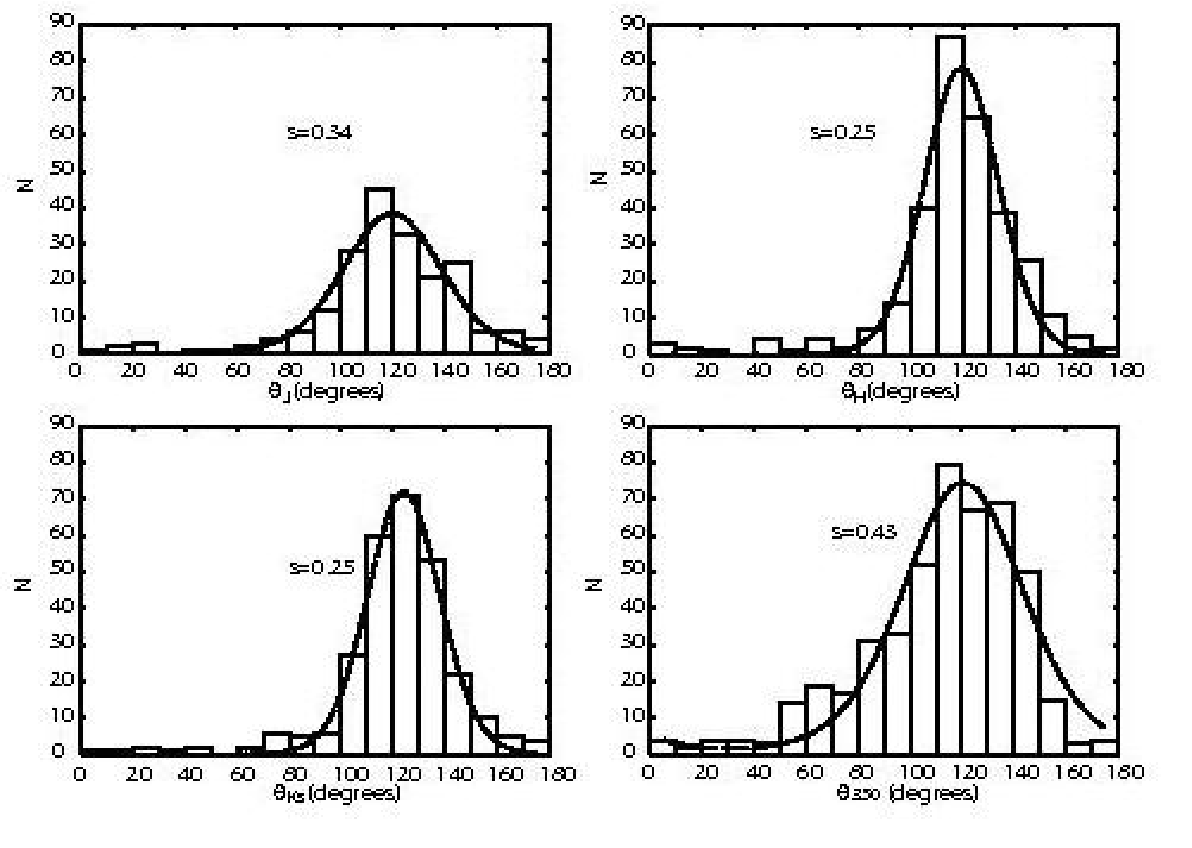}
\caption{Fitting of position angle distributions in a manner similar to
Myers \& Goodman (1991).}
\label{MG91}
\end{figure}

\clearpage

\begin{figure}[htb]
\includegraphics[width=15cm]{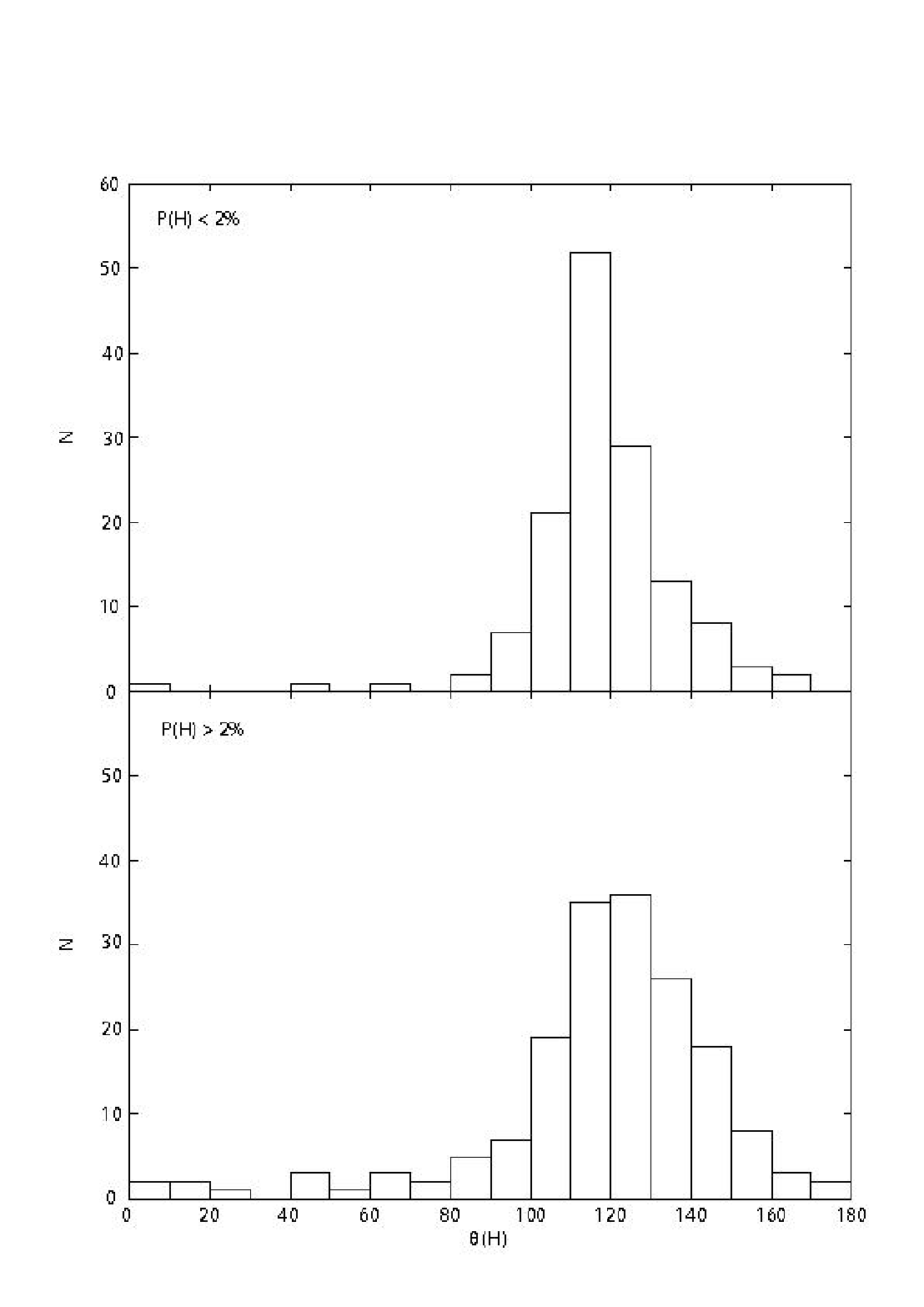}
\caption{
(Top) The histogram of 140 sources having polarizations lower than $2\%$.
(Bottom) The histogram of 173 sources having polarizations larger than $2\%$ .}
\label{2p}
\end{figure}

\begin{figure}[htb]
\includegraphics[width=15cm]{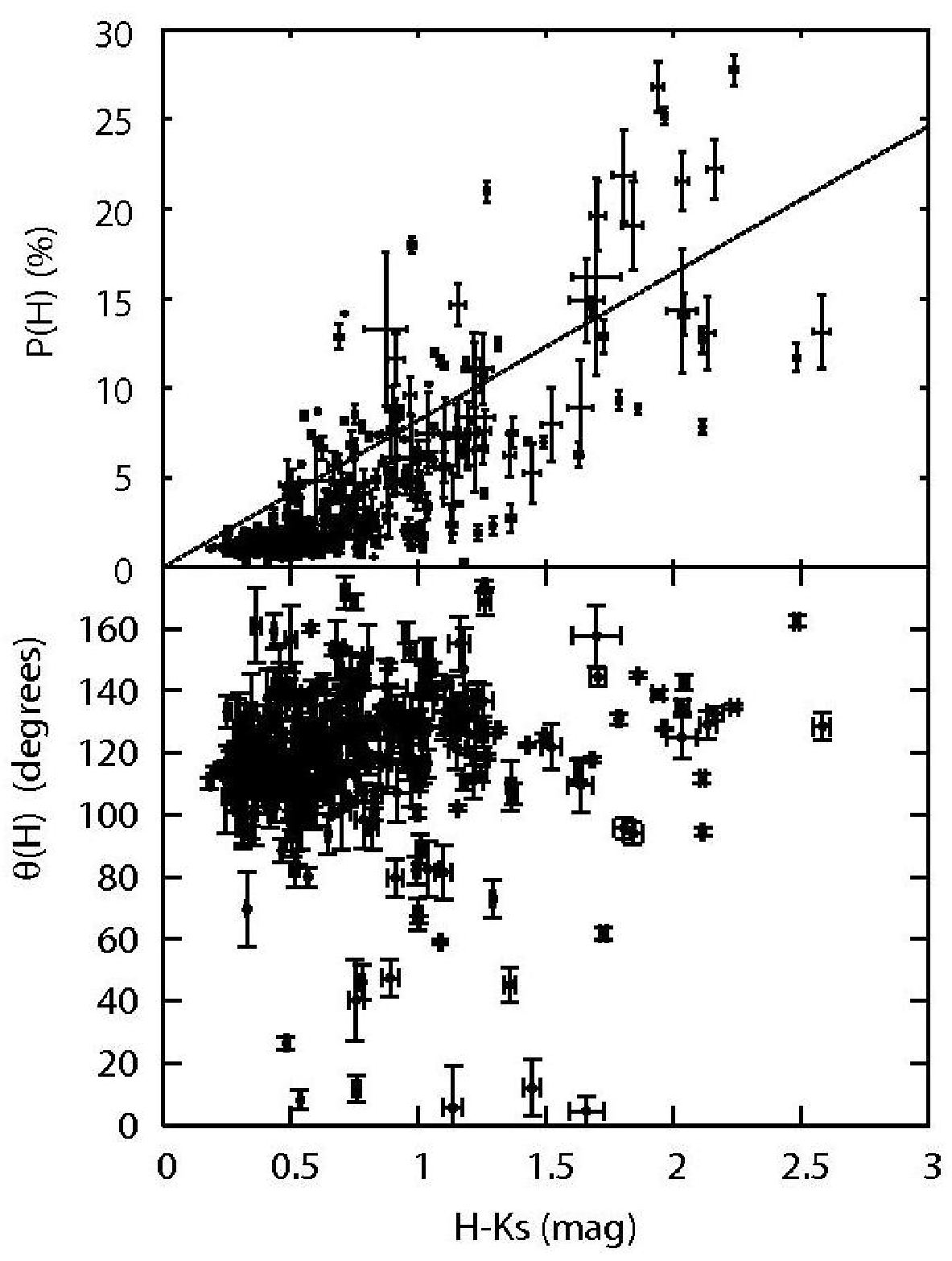}
\caption{
(Top) Degrees of polarization at $H$ vs. $H-Ks$ color.
The dashed line shows the assumed upper limit line of BN.
(Top) Degrees of polarization at $H$ vs. $H-Ks$ color.
(Bottom) Polarization position angles at $H$ vs. $H-Ks$ color.
}
\label{Pth_H-K}
\end{figure}

\clearpage
\begin{figure}[htb]
\includegraphics[width=15cm]{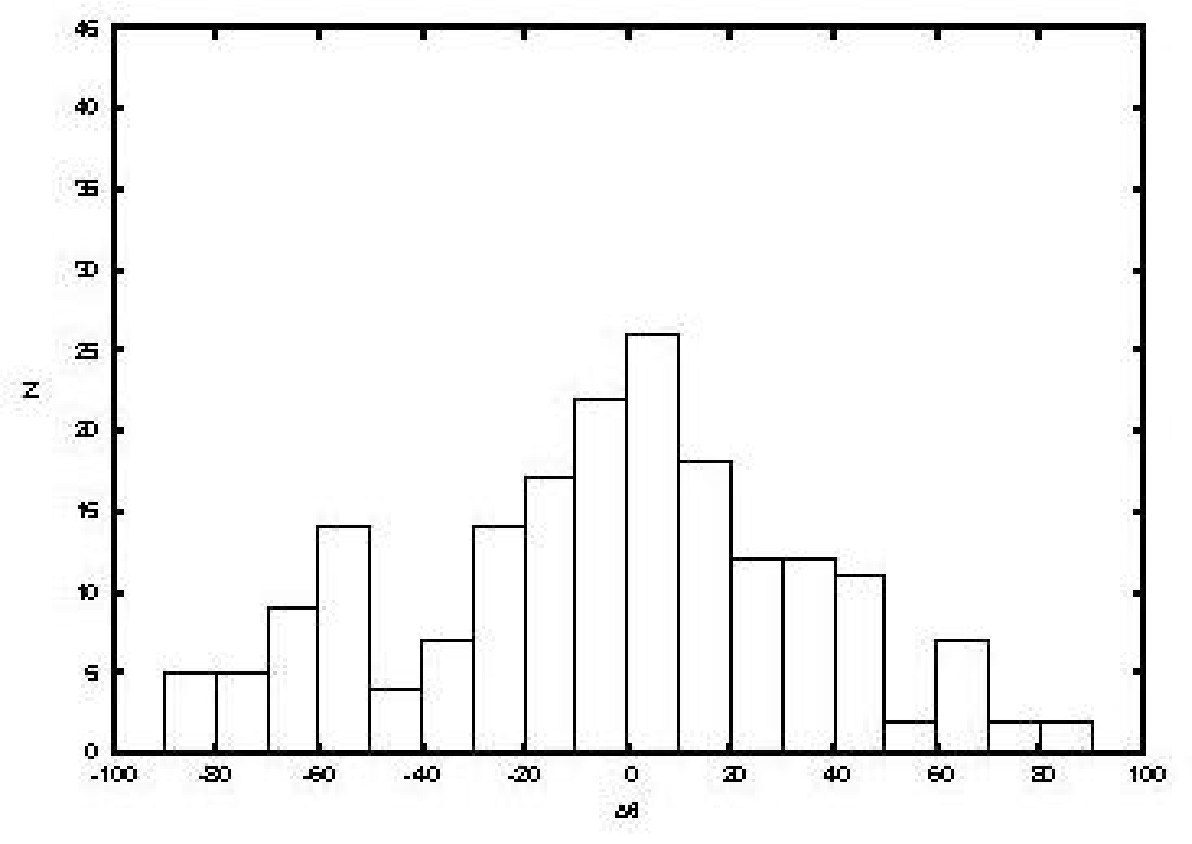}
\caption{Comparison of the sources having $P/(H-Ks) \leq P_{max}/(H-Ks)$ with submillimeter 
polarization position angles.}
\label{lowP}
\end{figure}

\clearpage
\begin{figure}[htb]
\includegraphics[width=10cm]{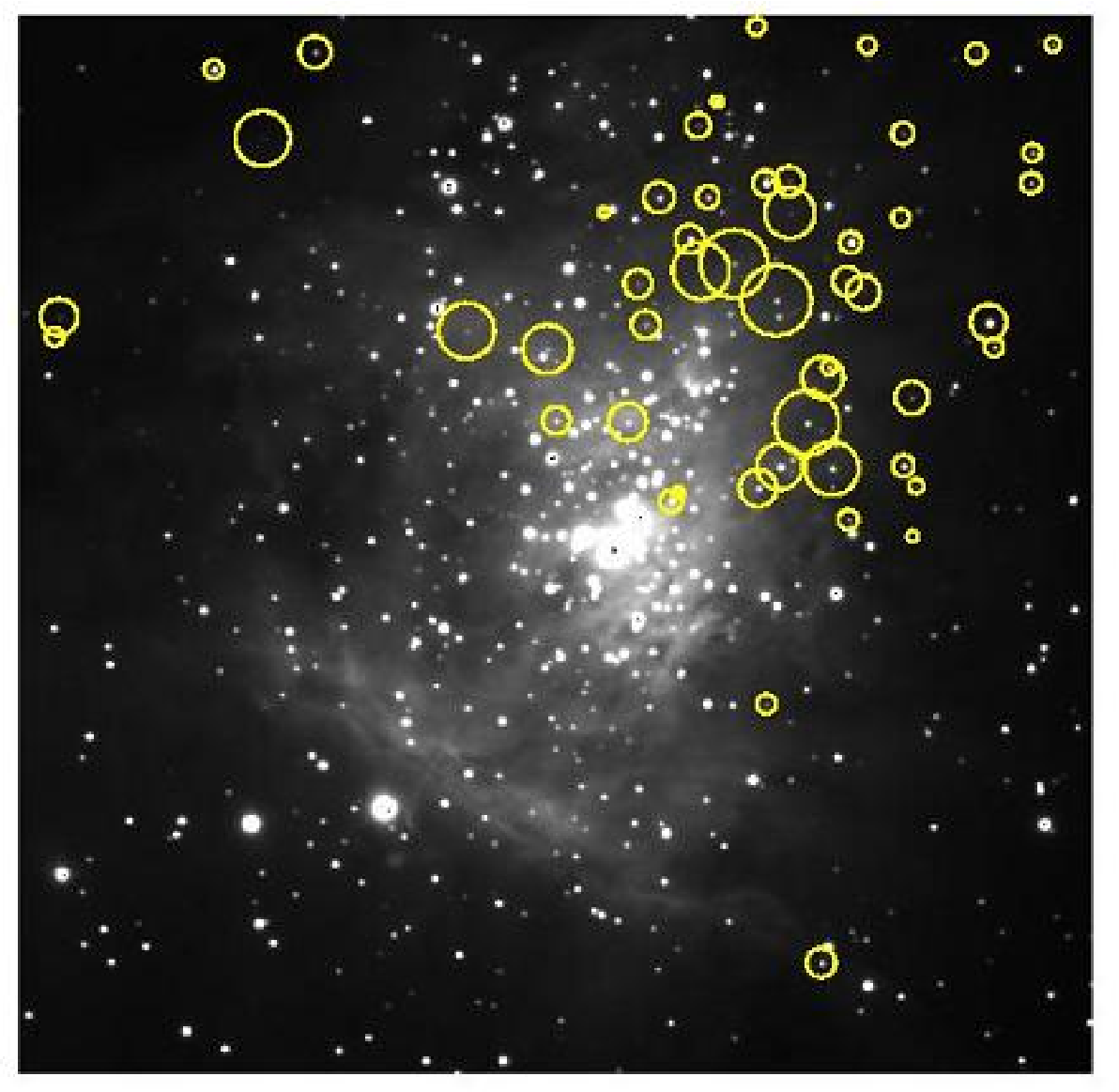}
\caption{Spatial distribution of highly polarized 51 sources.}
\label{Pmax_c}
\end{figure}

\clearpage
\begin{figure}[htb]
\includegraphics[width=15cm]{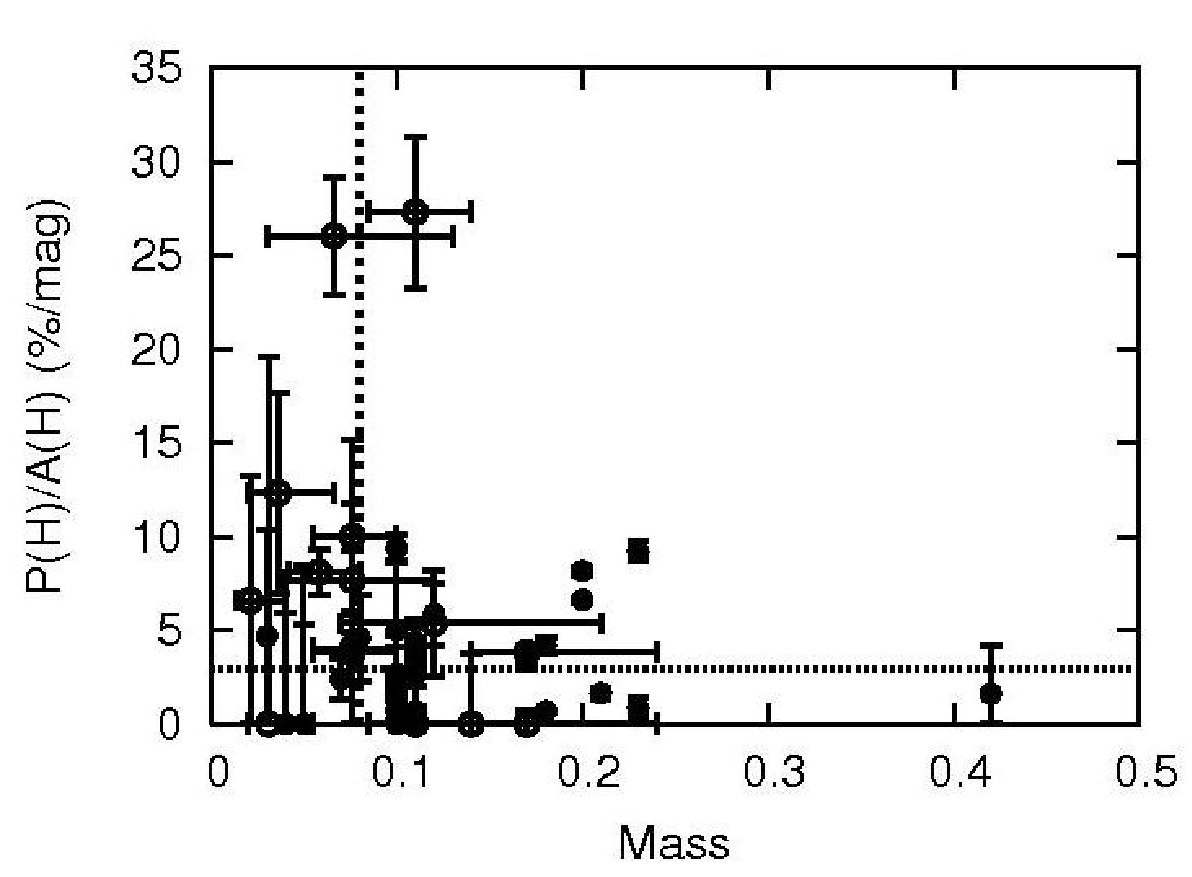}
\caption{$P(H)/A(H)$ for the 28 sources cross-identified with Slesnik et al. (2004) 
(filled circle) and Riddick et al. (2007) (open circle). 
The horizontal line indicates $P(H)/A(H)$=2.9, which corresponds to the assumed maximum value for interstellar polarization. The broken line indicates 0.08$M_\odot$, which corresponds to the brown dwarf mass limit.}
\label{mass}
\end{figure}



\end{document}